\documentclass[useAMS,usenatbib]{mn2e}

\usepackage{lscape}
\usepackage{graphicx}
\usepackage{amssymb}
\usepackage{multirow}

\title[Fitting the young main-sequence]{Fitting the young
main-sequence; distances, ages and age spreads.} \author[N.J.  Mayne
and Tim Naylor.]{N.J.  Mayne$^{1}$\thanks{E-mail:
nathan@astro.ex.ac.uk (NJM)} and Tim
Naylor$^{1}$\\
$^{1}$ School of Physics, University of Exeter, Stocker Road,
Exeter, EX4 4QL.\\}
\begin{document}

\date{Accepted ?. Received ?; in
  original form ?}

\pagerange{\pageref{firstpage}--\pageref{lastpage}} \pubyear{2007}

\maketitle

\label{firstpage}

\begin{abstract}
  We use several main-sequence models to derive distances (and extinctions),
  with statistically meaningful uncertainties for 11
  star-forming-regions and young clusters. The model dependency
  is shown to be small, allowing us
  to adopt the distances derived using one model. Using these
  distances we have revised the age order for some of the clusters of 
  \cite{2007MNRAS.375.1220M}. 
  The new nominal ages are: 
  $\approx2$ Myrs for NGC6530 and the ONC, 
  $\approx3$ Myrs for $\lambda$ Orionis, NGC2264 and $\sigma$ Orionis,
  $\approx4-5$ Myrs for NGC2362,
  $\approx13$ Myrs for h and $\chi$ Per,
  $\approx20$ Myrs for NGC1960 and 
  $\approx40$ Myrs for NGC2547. In cases of significantly variable extinction we
  have derived individual extinctions using a revised Q-method
  \citep{1953ApJ...117..313J}.  These new data show that the largest
  remaining uncertainty in deriving an age ordering (and
  necessarily ages) is metallicity. We also discuss the use
  of a feature we term the R-C gap overlap to provide a diagnostic of
  \textbf{isochronal} age spreads or varying accretion histories
  within a given star-formation-region. Finally, recent derivations of
  the distance to the ONC lie in two groups. Our new
  more precise distance of $391^{+12}_{-9}$ pc allows us to decisively
  reject the further distance, we adopt 400 pc as a convenient
  value.
  \end{abstract}


\section{Introduction}
\label{intro}

Colour-magnitude diagrams (CMDs) of star-formation regions (SFRs)
provide, in combination with model isochrones, an excellent tool with
which to determine distances, ages and individual stellar masses.
These parameters are critical for determining initial mass functions
(IMFs) for stellar populations and discovering the possible impacts of
local environment (such as the effect of ionising winds from massive
stars) on disc lifetimes and on star and planet formation and
evolution. Many calculations of IMFs, disc fractions etc are available
but they are derived in heterogeneous ways, thus hints of the effects
of environment are only recently beginning to emerge
\citep[e.g.][]{2007MNRAS.375.1220M,2004AJ....128..765S}.

Very precise photometry ($\approx1\%$) is routinely available,
along with sophisticated stellar models. However, current parameter
derivations from CMDs still have relatively large uncertainties and
are model dependent \citep[see the discussions
in][]{2004A&A...415..571B,
  2004ApJ...600..946P,2002MNRAS.335..291N,2007MNRAS.375.1220M}. Thus
current age (and distance) uncertainties all but `wash-out' any
environmental effects. Clearly more robust constraints would be
available for current stellar theories if more precise parameters
could be extracted from the CMDs of SFRs.

In \cite{2007MNRAS.375.1220M} we created an age ladder for
a range of pre-MS populations.
The first stage was to create empirical isochrones by fitting splines 
to the pre-MS locus.
Overlaying them in absolute magnitude and intrinsic colour results 
in an age ladder, with the youngest SFRs at the brightest absolute 
magnitudes.
SFRs with almost indistinguishable positions in the CMD were grouped,
and nominal ages assigned to each group.
Thus the age sequence (though not the nominal ages) are
free from the problems associated with pre-main-sequence (pre-MS) models. 
In \cite{2007MNRAS.375.1220M} we had to adopt literature distances for
the studied SFRs.  These distances were derived using a range of
different methods and their uncertainties proved to be the largest
remaining contributor to the uncertainties in our age ladder
placements.  Previous distances have chiefly been derived using
main-sequence (MS) isochrone fitting, pre-MS isochrone fitting or from
\textit{HIPPARCOS} parallax measurements. MS isochrone fitting
provides distances based on the positions of MS stars in a CMD, which
are independent of uncertainties in age. Pre-MS isochrone fitting also
uses the positions of stars in a CMD, but in this method the derived
distances are degenerate with age \citep[see
e.g.][]{2006MNRAS.373.1251N}. Finally, distances derived from
\textit{HIPPARCOS} parallax measurements are only available for a few
SFRs included in this paper, $\sigma$ Ori, NGC2547 and $\lambda$ Ori,
with all except $\lambda$ Ori having large uncertainties.

Of those methods used to derive distances, the most suitable for the
derivation of an age ladder is clearly MS isochrone fitting.  Fully
convective pre-MS stars (in young SFRs) are separated in a CMD from
those stars on the MS which have radiative cores. The transistion
region or gap in the CMD (measurable in colour), we term the
radiative-convective gap \citep[R-C gap, see][for introduction of
term and discussion]{2007MNRAS.375.1220M}. Once stars have crossed the
R-C gap their position in a CMD is almost independent of age until
they reach the turn off. However MS fitting has not yielded the
precision one would expect in distance estimates.  This is due to two
significant problems. Firstly the position of MS isochrones, although
temporally static, is model dependent, with different studies adopting
different models. Secondly distances are most often derived using `by
eye' fitting of models to the data, yielding ill-defined
uncertainties. A full discussion of previous fitting methods can be
found in \cite{2006MNRAS.373.1251N}.

In this paper we solve both of these problems. Firstly we show that
the model dependency is small for the model isochrones
studied.  We then adopt the distances from a single MS model.
This allows us to derive a set of precise distances,  accurate relative
to each other, which we term "relative distances". Second we use 
the $\tau^2$ fitting
technique \citep{2006MNRAS.373.1251N}, a new rigorous and
self-consistent method of fitting stars to isochrones, which yields statistically
meaningful uncertainties. This presents us with the opportunity to
achieve more precise distances from the fitting of high-mass (HM) or
MS stars.

The rest of this paper is laid out as follows. In Section \ref{data}
we detail the literature sources, the nature of the data used and any
sequence selection carried out on the stars.  Section
\ref{calibration} details the different model isochrones and
photometric calibrations used. Section \ref{fit_method} describes the
fitting process. This is done primarily by way of an example in
Section \ref{chiper_eg}.  Section \ref{ind_ext} describes the
derivation of individual extinctions, in particular Section \ref{Q}
describes an revised Q-method for calculating approximate individual
extinctions. The results for all the isochrone calibrations and
methods are presented in Section \ref{results_app}. Section
\ref{results} outlines our results for one adopted model with our
best-fitting distances given in Table \ref{dist_comp}. In section
\ref{implications} we discuss the implications of the individual
distances to several key SFRs (Section \ref{imp_ind}) and those of the
entire dataset.  
The implications of the dataset on metallicity
(Section \ref{imp_met}), age spreads and the R-C gap overlap (Section
\ref{imp_cuts}) and secular evolution within the SFRs with particular
reference to disc fractions (Section \ref{imp_sec}) are discussed.
The reader interested in distance and reddening
values should skip to Section \ref{results_app} for the values
derived from all the models used.

\section{The data}
\label{data}

All the datasets presented in this work are from literature sources.
To avoid distance-age degeneracy problems and to minimise the effect
of age assumption on our distance derivations we have only fitted HM
stars on or near the MS. We have used the memberships adopted
in the original source and made further photometric cuts. The sources
of photometric data and initial memberships are shown in Table
\ref{prep}. The photometry is in the Johnson-Cousins system
unless otherwise stated.

Further photometric cuts are required to remove non-member stars which
may have satisfied the membership criteria within the original
publication.  In addition photometric cuts are required to select the
correct part of the sequence. Motion in a CMD as a function of age is
rapid in the pre-MS and post-MS phases, compared to that on the MS.
Therefore inclusion of stars in these phases would introduce an age
dependency into our distances. 
Thus we make the two photometric cuts detailed below, which in all cases 
results in a clearly identifiable MS which lies clear of the contamination.

\subsection{The turn-off cut}
\label{seq_off}

If we include stars which are too bright they may have evolved away
from the MS (turn-off), so we make a photometric cut at the
turn-off, for the nominal age, despite the fact that in some cases the
MS appears to extend above the turn-off (see Section
\ref{implications} for a discussion of this discrepancy).

There is also a shift in the position of the MS as the isochronal age
increases, as the stars are `preparing' to turn-off. We have examined
this effect and the result of assuming the wrong age for the model
sequence. The effect is negligible in our experiment as explained in
Section \ref{fit_probs}.

\subsection{The turn-on cut}
\label{seq_on}

\begin{figure}
  \vspace*{174pt}
  \includegraphics{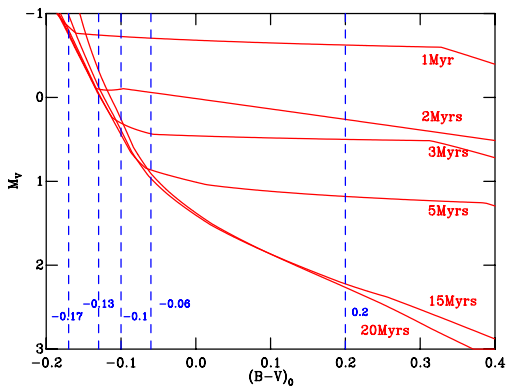}
\caption{1, 3, 5, 15 and 20 Myr isochrones from \citet{2000A&A...358..593S}
  showing the MS, and pre-MS. $B-V$ colour cuts are shown for each age
  to isolate the MS. These are: -0.17, -0.1, -0.06 and 0.2 for 1, 3, 
  5, and older sequences respectively.\label{ms}}
\end{figure}

If we include stars which are too red they may be pre-MS stars where 
age is degenerate with distance. 
The positions in colour at which stars join the MS (turn-on) and hence
where the turn-on cut should be made is predicted by pre-MS isochrones.
Figure \ref{ms} shows the pre-MS isochrones of
\cite{2000A&A...358..593S} for ages of 1, 3, 5, 15 and 40 Myrs (the
latter being our oldest SFR). The positions in colour of the turn-ons
are shown in Figure \ref{ms}. However for the younger SRFs the
observed MS in a CMD often appears to extend redder and fainter than
predicted by the pre-MS isochrones. So we have decided on
the positions of the cuts empirically, that is to say we have
identified the bottom of the MS in the data and placed the cut there,
at the blue edge of the radiative-convective gap. 
Table \ref{prep} shows the positions of
the cuts predicted from the isochrones of \cite{2000A&A...358..593S},
the reddening (see Sections \ref{extinction} and \ref{results}), the
actual cut employed and the age assigned in
\cite{2007MNRAS.375.1220M}. 
Using this emprical cut, as opposed to that from theory results in significantly 
more precise distances for some of our youngest SFRs.
This is because the distance is primarily derived from the curve of the
MS towards the red at fainter magnitudes.
A discussion of the implications of the MS extending below its
theoretical terminus can be found in Section \ref{implications}. 

\begin{table*}
\begin{tabular}{lcccccllc}
\multirow{2}{*}{SFR}&$B-V$&\multirow{2}{*}{$E(B-V)$}&\multicolumn{2}{|c|}{$(B-V)_0$
  Cut}&Nominal age (Myr)&Data&\multirow{2}{*}{Source}&Size\\
&Cut&&Data&Theory&\citet{2007MNRAS.375.1220M}&type&&(arcmins)\\
\hline
the ONC$^{(1)}$&-&-&0.1&-0.13&1&$VI$ \& $T_{eff}$&\cite{1997AJ....113.1733H}&20\\
NGC6530$^{(2)}$&0.5&0.32&0.18&-0.17&1&$UBVI$&\citet{2000AJ....120..333S}&15\\
NGC2244$^{(2)}$&0.45&0.47&-0.02&-0.13&$\approx2^{(3)}$&$UBVI$$H_{\alpha}$&\citet{2002AJ....123..892P}&15\\
NGC2264&-0.02&0.03&-0.05&-0.1&3&$UBVRI$&\citet{1980MNRAS.190..623M}&30\\
NGC2362&0.04&0.09&-0.05&-0.1&3&$UBV$&\citet{1953ApJ...117..313J}&30\\
$\lambda$ Ori&0.20&0.11&0.09&-0.1&3&$UBV$&\citet{1977MNRAS.181..657M}&30\\
$\sigma$ Ori&0.03&0.06&-0.03&-0.06&4-5&$V_TB_T$$^{(4)}$&\citet{Caballero.2007.}&30\\
$\chi$ Per$^{(7)}$&0.7&0.50&0.2&0.2&13&$UBV$&\citet{2002ApJ...576..880S}&2\\
h Per$^{(7)}$&0.74&0.54&0.2&0.2&13&$UBV$&\citet{2002ApJ...576..880S}&2\\
NGC1960&0.25&0.20&0.05&0.2&16$^{(5)}$&$UBV$&\citet{1953ApJ...117..313J}&10\\
NGC2547&0.1&0.04&0.06&$>0.2$&38$^{(6)}$&$UBV$&\citet{Claria.1982.}&15\\
\hline
\end{tabular}
\caption{The empirical and theoretical $B-V$ cuts for each SFR.
The apparent cut has been converted into intrinsic colour using
the $E(B-V)$ derived in Section \ref{results}.
Also shown is the nominal
age for each SFR from \citet{2007MNRAS.375.1220M}, the data type
used and source for the data and initial memberships. 
The notes are as follows.
(1) Individual extinctions derived from $T_{eff}$, see Section \ref{ind_ext}.
(2) Individual extinctions derived using the Q-method (see Section \ref{Q}).
(3) Age from \citet{2002AJ....123..892P}.
(4) Data in the $TYCHO$ photometric system.
(5) Age from \citet{2000A&A...357..471S}.
(6) Age from \citet{2006MNRAS.373.1251N}.
(7) For h and $\chi$ Per stars have additionally been
selected using apparent distance from the cluster centre \citep[as
defined in][]{2007MNRAS.375.1220M}, selecting a circular area around
the cluster centre of 2 arcmins in radius. \label{prep}} 
\end{table*}

\section{The models}
\label{calibration}

In this work we have used the MS stellar interior models of the Padova
\citep{2002A&A...391..195G} and Geneva \citep{2001A&A...366..538L}
groups. These provide an effective temperature ($T_{eff}$), luminosity
and surface gravity. These values must then be converted into colours
and magnitudes in the required photometric system
(Johnson-Cousins) to allow the fitting of photometric data.
Colours are found using a $T_{eff}$ to colour relation, and magnitudes
using the bolometric correction to the luminosity. Both the
colour-$T_{eff}$ relation and bolometric correction come from using
the parameters from a stellar interior model to find the correct model
atmosphere and then folding the resulting flux distribution through
appropriate photometric filter responses.  Once this is achieved the
photometric colours and magnitudes must then be calibrated to a
standard scale, using the colours of Vega in the photometric system.
We have used three main isochrone and extinction systems calibrated to
two different Vega colour systems, and we now detail each.

\subsection{Geneva}
\label{geneva}

The Geneva isochrones \citep[as provided in][]{2001A&A...366..538L}
are from the Geneva stellar interior models (basic set) in conjunction
with the updated BaSeL-2.2 model atmospheres from
\cite{1999ASPC..192..203W}. To derive photometric magnitudes ($V$)
they have adopted the bolometric corrections from
\cite{1998A&AS..130...65L} which are defined to fit the empirical
scale of \cite{1996ApJ...469..355F} (not calibrated to the Sun).
They calculate colours for Johnson-Cousins photometry
using the filter response functions of \cite{1978A&A....70..555B}
($UBV$) and \cite{1979PASP...91..589B} ($RI$). The colours and
magnitudes of the isochrone are then calibrated to Vega colours of
zero.  For these isochrones we use the canonical
extinction vectors, namely $E(U-B)/E(B-V)=0.73$, $A(V)/E(B-V)=3.1$ and
$A(V)/E(V-I)=0.41$.

\subsection{Geneva-Bessell}
\label{geneva_bes}

For the Geneva-Bessell isochrones we have used the interior models of
\cite{2001A&A...366..538L}, specifically their basic model set (``c'')
generally applicable for stars with $M<12M_{\odot}$. Their conversion
to photometric colours follows that of \cite{1998A&A...333..231B}.
\cite{1998A&A...333..231B} use the ATLAS9 atmosphere models of
\cite{1997A&A...318..841C} (at solar metallicity only) and the filter
responses of \cite{1990PASP..102.1181B} ($UBVRI$), to calculate the
colour-$T_{eff}$ relations and bolometric corrections. The resulting
Johnson-Cousins photometry is then calibrated to Vega colours
of zero. In practice we carry out these conversions in our own code,
since that allows us to also calibrate to what we term the non-zero
system where $(B-V)_{Vega}=-0.002$ and $(U-B)_{Vega}=-0.004$. 

\subsection{Padova-Bessell}
\label{Padova_bes}

For the Padova-Bessell isochrones we use the stellar interior models
of \cite{2002A&A...391..195G}. The colours and magnitudes are then
calculated using the conversions of \cite{1998A&A...333..231B} as for
the Geneva-Bessell isochrones. The resulting Johnson-Cousins
photometry is then calibrated to either Vega colours of zero or the
non-zero system. 

\subsection{Extinction vectors in the Bessell system}
\label{ext_vect}

It is well known that extinction vectors are actually a function of
intrinsic colour (or spectral type).
\cite{1998A&A...333..231B} provide extinction vectors as a function of
colour based on the extinction curves of \cite{1990ARA&A..28...37M}.
We therefore use these extinction vectors for the Geneva-Bessell and
Padova-Bessell models.

The extinctions are provided for an $E(B-V)=0.3$. 
To check the range of extinctions to the SFRs studied here does not have 
a significant effect, we folded the
solar abundance ATLAS 9 spectra with the ``new'' opacity distribution
function \citep{2004astro.ph..5087C} through the bandpasses of 
\cite{1998A&A...333..231B}.
We then reddened the spectra
according to the prescription of \cite{1989ApJ...345..245C} to yield
an $E(B-V)$ of approximately 0.3. (This function is the one tabulated
in the \cite{1990ARA&A..28...37M} paper used by
\cite{1998A&A...333..231B}).
We used an $R_v$ of 3.2 since, when folded through the bandpasses of 
\cite{1990PASP..102.1181B} we found this gave the best match to the $BV$ extinction 
vector of \cite{1998A&A...333..231B}.
We calculated the difference in $A_v/E(B-V)$ for values of $E(B-V)$ of 1 
(typical of the SFRs we have fitted) and 3 (approximating to the highest 
reddening of any SFR fitted).
For $B-V<1.5$ we find the largest difference is -0.02 mags, which has 
a negligible impact on our fits.

\subsection{$TYCHO$ photometry}
\label{tycho}

For $\sigma$ Orionis the data were taken in the $TYCHO$ photometric
system. We have fitted these data using the conversion of
\cite{2000PASP..112..961B} transform our Geneva-Bessell isochrones
into the $TYCHO$ system. We have defined extinction vectors in the
$TYCHO$ photometric system, in a process similar to that described in
Section \ref{ext_vect}.
We set our zero-points by requiring we reproduced the $T_{eff}$ vs
$B-V$ relationship of \cite{1990PASP..102.1181B} and the $B-V$ vs $\Delta(B-V)$ 
relationship of \cite{2000PASP..112..961B}.
This gave, for $(B-V)_T<0.065$
\begin{equation}
{{(A_V)_T}\over{E(B-V)_T}}=3.358+0.237(B-V)_T,
\end{equation}
and for $0.065<(B-V)_T<0.5$
\begin{equation}
{{(A_V)_T}\over{E(B-V)_T}}=3.387-0.207(B-V)_T.
\end{equation}
These fits never deviate from the calculated curve by more than about 0.01
mags.
At $B-V$=0 the $BV$ curves match to within 0.01 mags, though this worsens
to 0.05 mags at $B-V$=1.

\section{The fitting method}
\label{fit_method}

Throughout this example and later sections all the isochrone fits
displayed and tests undertaken use the Geneva-Bessell Vega-zero
isochrones.  These isochrones are also adopted in Section
\ref{results} where we draw implications from the resulting distances
and age ordering.  As shown in Section \ref{results_app} the model
dependency between the different isochrones is practically very small.

\subsection{Example fit: $\chi$ Per}
\label{chiper_eg}

It is most instructive to describe the fitting method via an example.
Here we use the cluster $\chi$ Per which is approximately 13 Myrs old
\citep{2007MNRAS.375.1220M}, at a distance modulus of $11.85$
\citep{2002ApJ...576..880S} with an $A_V\approx1.6$
\citep{2007MNRAS.375.1220M} (the extinction is reasonably uniform).
First, we describe the derivation of distance.  Deriving a distance
does require a known extinction, the derivation of which is described
later in this section.

\subsubsection{Fitting statistic, distance derivation}
\label{fitting_stat}

The fitting statistic used in this work is $\tau^2$, which is
introduced in \cite{2006MNRAS.373.1251N}. Fitting of MS data using
this technique is described in \cite{2007MNRAS.376..580J}. $\tau^2$ is
essentially a generalised $\chi^2$ statistic including uncertainties
in two dimensions, and models with a two-dimensional distribution as
opposed to a single isochronal line.  The best fitting model is found
by minimising $\tau^2$.

Once we have selected an isochrone we use a Monte-Carlo method to
generate a colour-magnitude probability grid. Each pixel in this grid
is assigned a value that gives the probability of finding a star drawn
from the population represented by the isochrone, at any given colour
and magnitude. The model is then adjusted through a range of distances
and the values of $\tau^2$ for each star summed to calculate a total
value of $\tau^2$ for each distance step, as detailed in
\cite{2006MNRAS.373.1251N}. These $\tau^2$ contributions were
clipped, i.e.  the contribution to the total $\tau^2$ for any single
data point value is capped at some set value. Clipping avoids
erroneously included non-members or anomalous objects many $\sigma$
from a given model overwhelming the result. The lowest total $\tau^2$
was then selected as the best fitting model.

Once a given fit was completed a probability of obtaining the
resulting $\tau^2$, $P_r(\tau^2)$, was calculated.  If $P_r(\tau^2)$
is far from 50\%, the fitting is repeated with an additional
systematic uncertainty added to the data. This is analogous to
enlarging error bars to achieve a $\chi^2_\nu=1$. This process ensures
the model is a good fit to the data and we are then able to derive the
parameter uncertainties. To derive these uncertainties, we used the
bootstrap method described in \cite{2006MNRAS.373.1251N}, repeating
the fitting 100 times and deriving 68\% confidence intervals.

\begin{figure}
  \vspace*{174pt}
  \includegraphics{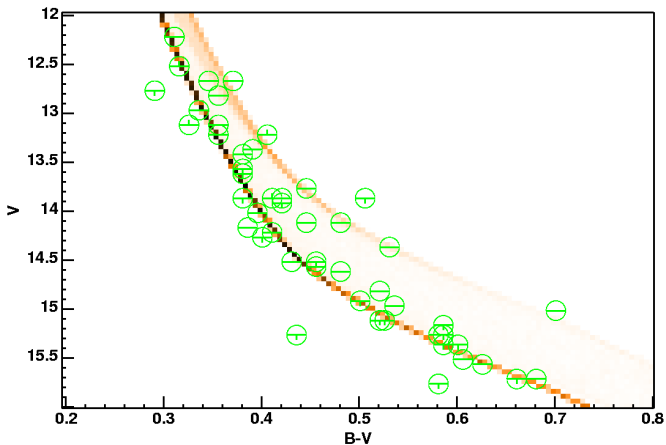}
\caption{The distance fit for $\chi$ Per to the Geneva-Bessell
  isochrones. See Section \ref{chiper_eg} for details of the symbols.
  \label{chiper_fit}}
\end{figure}

The resulting fit for $\chi$ Per is shown in Figure \ref{chiper_fit}.
The derived distance modulus and 68\% confidence interval is
$11.79<11.83<11.88$.  This agrees with the most recent literature
derivation of $11.85\pm0.05$ from \cite{2002ApJ...576..880S}, which
was the same dataset.

Figure \ref{chiper_fit} and all subsequent figures showing fitted data
have several elements requiring explanation. The shaded area shows the
probability density of finding a star at a particular colour and
magnitude (or colour and colour for extinction fitting).  This
density, $\rho$ is that from Equation 1 of \cite{2006MNRAS.373.1251N}
generated for a specific isochrone, in this case Geneva-Bessell. The
circles show the positions of the photometry and the bars give the
uncertainties in magnitude and colour. For all the figures showing
fitted data (except those using individual extinctions, see Section
\ref{ind_ext}) the models have been adjusted to the natural space of
the data i.e. apparent colour and magnitude.
 
\subsubsection{Mean extinction fitting}
\label{extinction}

\begin{figure}
  \vspace*{174pt}
  \includegraphics{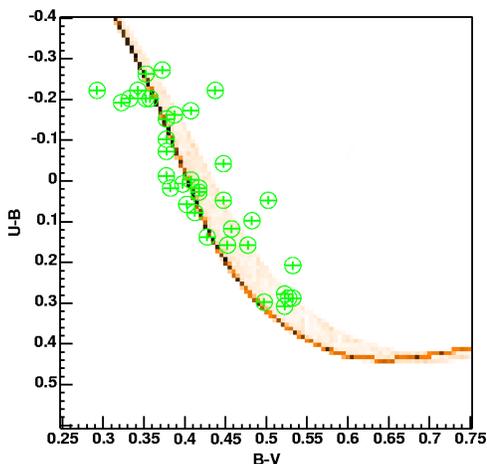}
\caption{Figure showing $E(B-V)$ fit of $\chi$ Per to the Geneva-Bessell
  isochrones.  See Section \ref{extinction} for details of the
  symbols. Here a systematic shift in $E(B-V)$ as a function of colour
  ($U-B$) is evident, as in Figure \ref{hper_fitebv}. This is probably
  due to a difference in the photometric systems of
  \citet{1998A&A...333..231B} and \citet{2002ApJ...576..880S} (see
  Section \ref{Q}). \label{chiper_fitebv}}
\end{figure}

To allow us to derive a distance, an extinction is required and is
indeed crucial as changing it will change the distance derived for the
stars we are fitting. We can derive a mean extinction by fitting the
data to an isochrone in a colour-colour diagram. Where we have $UBV$
photometry we have simply fitted the sequence in $U-B$ vs $B-V$ in a
similar fashion as that for a distance. However, instead of changing
the distance, we evaluate $\tau^2$ at different values of the
reddening. The resulting fit can be seen in Figure
\ref{chiper_fitebv}, with a best fitting $E(B-V)=0.50$. Figure
\ref{chiper_fitebv} and all the subsequent figures showing extinction
fitting contain the same components as those for distance fitting.

In general fitting for an extinction using the $\tau^2$ method is
desirable as photometric individual extinction methods rely on the
star being a sinlge star or an equal mass-mass binary (see in Section
\ref{Q}). However, in some cases the dispersion in an $E(B-V)$ fit is
too large, i.e. the scatter around the isochrone in the $U-B$ vs $B-V$
is too large to confidently assign one mean extinction.  This is the
case where there is significantly variable reddening across a SFR.
Here we are forced to derive reddenings and therefore extinctions for
each star from $UBV$ photometry alone using the Q-method
\citep{1953ApJ...117..313J}. This is the case for NGC6530, NGC2244 and
$\lambda$ Orionis. In addition in the case of the ONC $UBV$ colours
are not available, here alternative methods must be used to derive
extinctions.  All these individual extinction derivations are detailed
in Section \ref{ind_ext}.

\subsection{Practical effect of assumptions}
\label{fit_probs}

During fitting we have made three main assumptions about the MS and it
is important to examine the validity of these.  Firstly we have used
an isochrone at an assumed nominal age which could be inaccurate for a
given SFR. This may be important both just prior to turn-off and at
the turn-on. Secondly, by using MS isochrones we have implicitly
assumed that none of the stars fitted are on the pre-MS.  Lastly, we
have used isochrones of approximately solar composition
($Z=0.02$).

\subsubsection{Age assumption}
\label{age_effects}

As a (coeval) stellar population ages, stars of a decreasing mass
turn-off from the MS. In addition, stars close to but nominally below
the turn-off age also move slightly red-wards in position in a CMD. As
we remove stars brighter than the turn-off (Section \ref{seq_off})
we avoid the age dependency of this feature, but we still retain the
age dependency of the upper-MS. This shift of the upper-MS is small
and furthermore in principle is modeled as we use a MS at the nominal
age. However it is important to quantify this effect to ensure our
distances are robust against assuming an incorrect age.

To test this upper-MS age dependency we have simulated photometry of a
population based on a 1Myr Geneva-Bessell isochrone. We have then
fitted these data (as described in Section \ref{chiper_eg}) across a
typical colour range, specifically blue-ward of $(B-V)_0=0.2$ (as
shown in Table \ref{prep} cuts in all our SFRs are blue-ward of this),
to a 10Myr Geneva-Bessell isochrone.  This provides a measure of any
extra uncertainty one accrues if the age assumption is incorrect. The
resulting distance modulus and reddening are $dm\approx0.001$
and $E(B-V)\approx-0.028$, for this factor ten in age. The effect of
assuming an incorrect age on a derived distance and $E(B-V)$ is
therefore negligible even for a large error in assumed age.

\subsubsection{MS isochrones for a possible PMS population}
\label{pms_not_ms}

As pre-MS stars approach the MS, just prior to the onset of hydrogen
burning they enter a quasi-equilibrium state as hydrogen ignition
starts. This delays their arrival onto the MS.  The rate at which this
phase progresses is a function of stellar mass and is, over the mass
range of interest, very short in comparison to the evolution of the
pre-MS.  However, as such stars are slightly brighter than MS
magnitudes, the distances derived may be systematically reduced as a
function of age.

Some stars in our selected sequence may still be in this pre-MS phase
even though appearing to be on the MS. This effect is greatest for the
younger SFRs. However within the colour cuts we have used the
deviation from the MS is a maximum of $\approx0.05$ mag (in magnitude)
and only affects a limited number of stars. Thus its overall effect
will be much smaller than 0.05 mags.

\subsubsection{Composition}
\label{composition}

As compositions are not available for all the SFRs studied we have
assumed solar metallicity ($Z=0.02$). However there is evidence to
suggest that some SFRs have a metallicity as low as half-solar (see
Section \ref{imp_met}). We must therefore quantify the possible effect
varying composition could have on our relative distances. To do this
we have again simulated a population, at 10Myr, using the Geneva
isochrones with $Z=0.008$ (the closest match in the library to
half-solar) and fitted these data for extinction and distance with
solar composition isochrones (as described in Section
\ref{chiper_eg}).  The resulting differences are, for reddening
$E(B-V)=0.005$ and for distance modulus, $dm\approx0.41$. This is a
negligible difference in reddening and therefore extinction but a
significant error in distance modulus.  We have also fitted $\chi$ Per
using the half-solar metallicity isochrones. Here we must use the
colours and photometric system provided with the Geneva isochrones as
our Geneva-Bessell isochrones utilise atmospheres of solar
metallicity. The resulting distance modulus is $11.32<11.36<11.41$ and
a reddening of $E(B-V)=0.51$ (using uncertainties of 0.018 for a
$P_r(\tau^2)\approx0.5$). This means that if the metallicity is indeed
$Z\approx0.01$ the distance modulus to $\chi$ Per is actually
$\approx0.5$ mags less, making the stars older.  This could have a
major effect on the distances and ages of star-forming regions.
Therefore composition information is vital in the future for more
accurate SFR parameters. A further discussion of this problem can be
found in Section \ref{implications}.

\section{Individual extinctions}
\label{ind_ext}

As stated in Section \ref{extinction} it is sometimes not possible or
desirable to derive a mean extinction using the $\tau^2$ fitting
method. In this Section we detail the cases where individual
extinctions have been derived. Source by source extinctions have been
used for six SFRs, using two methods. In all of these cases the
extinctions applied will change the relative positions of the stars in
colour magnitude space, as opposed to applying a mean extinction where
the entire dataset is simply translated (shifted). Therefore, in the
case of individual extinctions these must be applied prior to fitting,
and the resulting figures show the data and model in intrinsic colour
and extinction free magnitude.

\subsection{Effective temperatures}
\label{int_col}

\subsubsection{the ONC}
\label{int_onc}

For the ONC $UBV$ photometry is unavailable, so fitting for a mean
extinction using the method detailed in Section \ref{extinction} is
not possible. However \cite{1997AJ....113.1733H} provides $VI$
photometry an of the centre of the ONC. We have used
these temperatures, assumed a surface gravity ($log(g)$)  of 4.5 and
interpolated to obtain intrinsic colours from the $T_{eff}$ relations
in \cite{1998A&A...333..231B}.  This allows us to derive an $A_V$ from
$E(V-I)$ using $A_V=((3.26+0.22(B-V)_0)/(1.32+0.06(V-I)_0))\times
E(V-I)$ \citep[derived from the extinction vectors
of][]{1998A&A...333..231B} and therefore calculate an unredenned
magnitude.  These resulting magnitudes and colours can then be fitted
to derive a distance as in Section \ref{chiper_eg}.  For the analysis
in Section \ref{results} of this paper we have fitted the data using
the intrinsic $B-V$ values appropriate for the effective temperatures, 
but we have also fitted the $V-I$ data, the
result of which can be found in Section \ref{results_app}.

\subsubsection{h and $\chi$ Per}
\label{int_per}

The check the veracity of this technique, we compared the distances
it yields for h and $\chi$ Per, with those from Sections \ref{fit_method} 
and \ref{results_app}.
We used the $T_{eff}$ data from \cite{2002ApJ...576..880S}
to derive intrinsic colours. However, as the $T_{eff}$ information is
only available for the hottest stars, the colour range for fitting is
small.  Practically this means the uncertainties are large,
particularly towards greater distances, but the answers for
both techniques are consistent.
As a further test,  we have supplemented the objects with $T_{eff}$ data,
with stars from the main
catalogue dereddened to intrinsic colours using the mean extinction
from the $\tau^2$ fitting.
All the two techniques give consistent distances (Table \ref{distances_int_2}), but those using the 
combined photometric and $T_{eff}$ datasets are more precise
(ranges of 0.01 and 0.04 mags for h and $\chi$ Per respectively).
This confirms that the $T_{eff}$ technique gives consistent answers, but
we are wary of adopting the more precise technique for the remainder of this paper
since our primary aim 
is to derive distances for many SFRs using a single method.

\subsection{Q-method}
\label{Q}

\begin{figure}
  \vspace*{174pt}
  \includegraphics{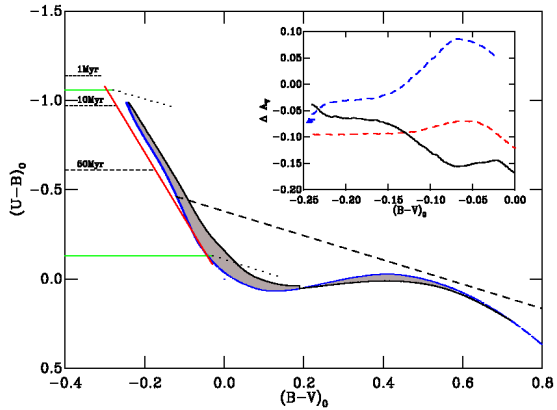}
\caption{Geneva-Bessell 10 Myr isochrone including binaries 
  (gray region enclosed by black line) and the Q-method MS straight
  line (red line). The horizontal dashed lines are the points at which
  the Geneva-Bessell isochrones evolve away from the MS.  The
  horizontal bold lines are the validity region of the original
  Q-method. The two angled dotted lines are the extinction vectors at
  each end of the Q-method validity range from
  \citet{1998A&A...333..231B}. Finally the large dashed line shows the
  region below which the solution for a given star can be
  multi-valued. The inset shows the differences in derived extinction
  ($\Delta A_V$).  The top dashed line shows $\Delta A_V$ in the sense
  of the revised Q-method minus the old Q-method.  The lower dashed
  shows $\Delta A_V$ derived from the reddest binaries minus the old
  Q-method.  The bold line is $\Delta A_V$ derived from the reddest
  binaries minus the single star sequence.\label{q_prob}}
\end{figure}

The Q-method is a widely used method to derive individual
source-by-source reddenings and therefore extinctions using a $U-B$ vs
$B-V$ colour-colour diagram.  \cite{1953ApJ...117..313J} model the MS
as the straight line in Figure \ref{q_prob}.  As this Figure shows,
this can result in errors in the derived $A_V$ of up to 0.1 mag, to
which must be added further 0.08 mag due to using colour independent
extinction vectors.  These are not intrinsic failings of the method,
fitting a straight line to a modern isochrone, and using
colour-dependent extinction vectors yields
\begin{equation}
Q=0.24(B-V)^2+3.257(B-V)+0.015,
\label{1}
\end{equation}
in the \cite{1998A&A...333..231B} system.  Within the ranges of colour
in Table \ref{q_lim}, this only differs by $<0.005$ mag from
interpolating onto the isochrone.  However in our ``revised Q-method" we
derive reddenings using isochone interpolation.

\begin{table}
\begin{tabular}{|l|c|c|}
Age&$U-B$&$B-V$\\
\hline
1&$-0.17<U-B<-1.15$&$-0.06<B-V<-0.30$\\
3&$-0.17<U-B<-1.12$&$-0.06<B-V<-0.29$\\
5&$-0.17<U-B<-1.08$&$-0.06<B-V<-0.28$\\
15&$-0.17<U-B<-0.90$&$-0.06<B-V<-0.23$\\
30&$-0.17<U-B<-0.72$&$-0.06<B-V<-0.22$\\
\hline
\end{tabular}
\caption{The approximate validity range for Equation \ref{1} (not including
  multiple valued solutions) for several ages derived from the
  Geneva-Bessell isochrones, defined as regions where the real
  isochrone moves more than 0.01 in colour away from the straight line
  model. \label{q_lim}}
\end{table}

Implicit in using a model, are assumptions as to the age and
metallicity of the isochrone.  We have tested the effect of both of
these, and provided one remains within the colour ranges given in
Table \ref{q_lim}, their effect on the derived $A_V$ is less than
$0.005$ mags.  The blue limit in Table \ref{q_lim} corresponds to 
the turn-off, the red limit to
approximately $U-B=B-V=0$.  A further possible concern is shown in
Figure \ref{q_prob}; once the intrinsic colour is redward of
$U-B\approx$-0.2, there is the possibility that the star actually
lies on a redder part of the isochrone, leading to an ambiguity in the
extinction.  However, if the object is on the wrong part of the
isochrone, the extinction is clearly anomalous, provided that the
scatter in extinctions between stars is small.

The major concern when calculating individual reddenings and therefore
extinctions using a colour-colour diagram is the effect of binarity.
Figure \ref{q_prob} shows the range of colours occupied by
unequal-mass binaries.  In the absence of multiplicity information,
any colour-colour method must dereden stars onto the single
star/equal-mass binary sequence (see Section \ref{extinction}).  As
can be seen in Figure \ref{chiper_fitebv}, without multiplicity
information it is unclear whether a star should lie on the single
star/equal-mass binary sequence or in the region occupied by
unequal-mass binaries.  The effect and range of this problem is shown
in Figure \ref{q_prob}.  The outer binary envelope has been modeled
and the possible differences in derived $A_V$ found.  Figure
\ref{q_prob} shows that binarity has an effect of up to $\Delta
A_V\approx0.15$ on the derived extinction.  This effect becomes
increasingly significant as one moves down (redder in $U-B$) the MS
isochrone.

As the revised Q-method does not account for the scatter
binaries produce in a CMD, or colour-colour diagram, it is a
statistically ill-defined process. 
Effectively most of the intrinsic scatter from the binary sequence and
photometric uncertainties is removed, in addition to that caused by
variable extinction.  
Therefore, only when there is good evidence, from the mean extinction
fitting method, that the extinction in a SFR is large and variable do
we apply the revised Q-method.
Thus we formulate the null hypothesis that the reddening or extinction
is uniform.
We fit to derive a mean extinction and subsequently
a distance.  
These results can be found in Table \ref{distances_UBV} and are
discussed in Section \ref{results_app}.
However, in some cases of distance fitting, after applying a mean
extinction, the addition of large systematic uncertainties was
required to return a $Pr(\tau^2)\approx50\%$.
In these cases we are forced to reject the null hypothesis and use the
revised Q-method to derive individual extinctions.  
We believe that additional systematic uncertainties of up to 2\% are
credible, therefore we apply the revised Q-method in cases where our
added systematic uncertainties exceed this level.

\subsection{SFRs with significantly variable reddening}
\label{q_eg}

\begin{figure}
  \vspace*{174pt}
  \includegraphics{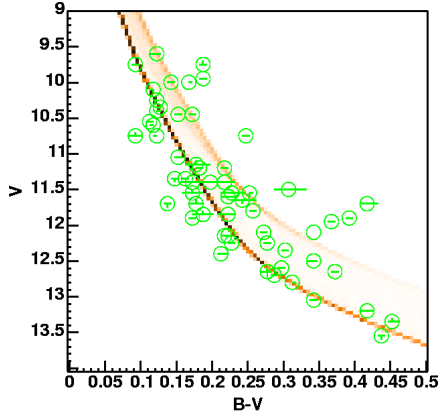}
\caption{NGC6530 distance fit using mean extinction, the fit for which
  is shown as Figure \ref{ngc6530_fitebv}.\label{ngc6530_fit}}
\end{figure}

\begin{figure}
  \vspace*{174pt}
  \includegraphics{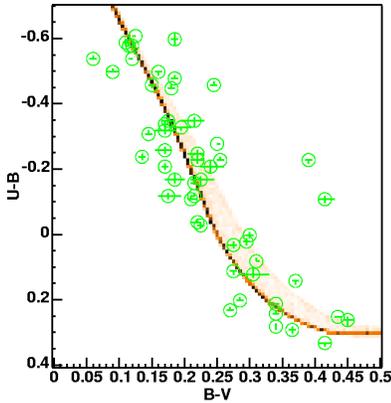}
\caption{NGC6530 $E(B-V)$ fit.\label{ngc6530_fitebv}}
\end{figure}

\begin{figure}
  \vspace*{174pt}
  \includegraphics{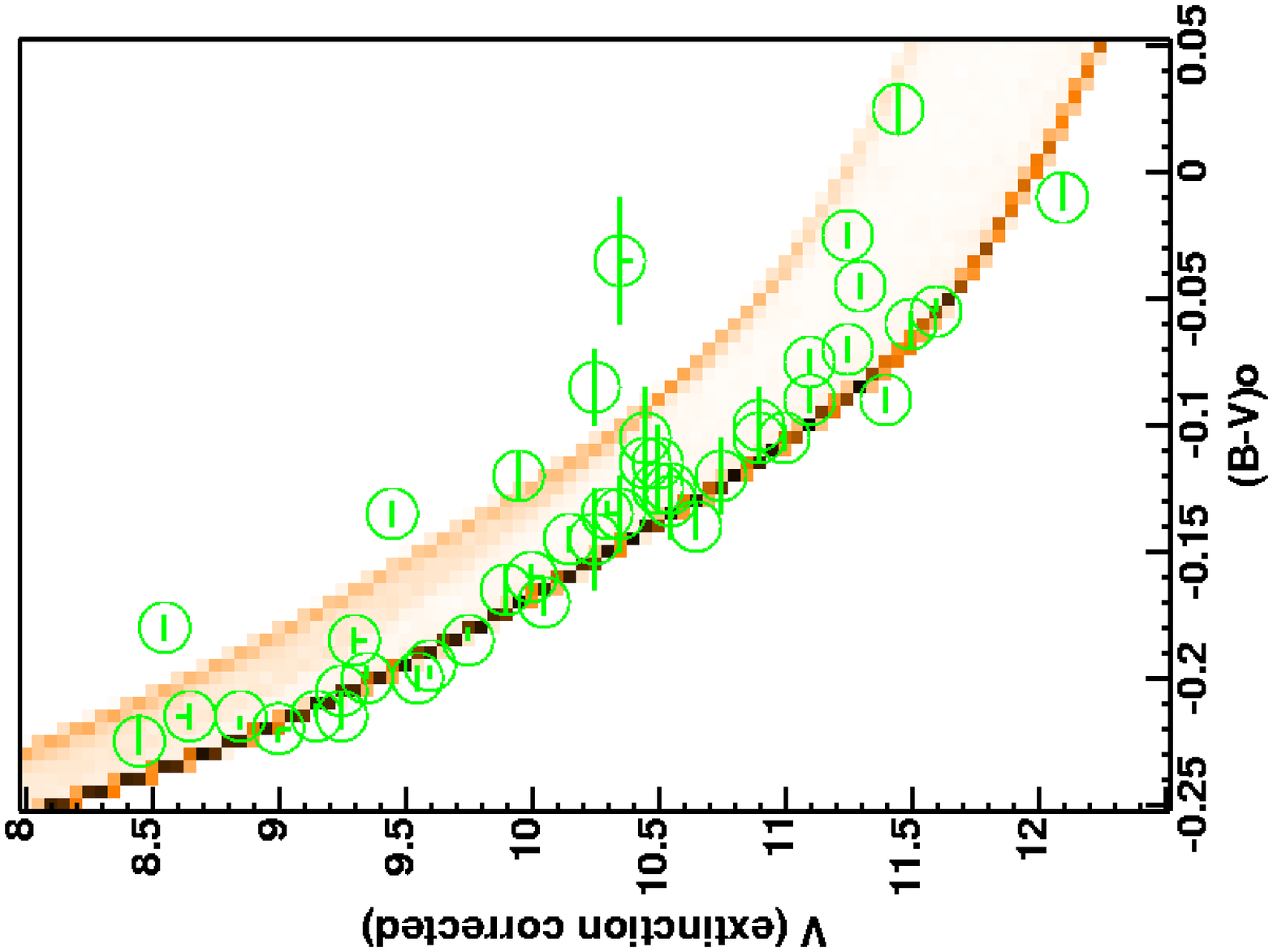}
\caption{NGC6530 distance fit using Q-method extinctions (see Section
  \ref{Q}). Figure shown in intrinsic colour and extinction corrected
  magnitude (see Section \ref{ind_ext}).\label{ngc6530_qfit}}
\end{figure}

In Figure \ref{ngc6530_fitebv} we show the $\tau^2$ mean extinction fit
for NGC6530.
There is clearly a large scatter, which is also reflected in the corresponding
distance fit, Figure \ref{ngc6530_fit}.
This scatter improves significantly when we use the revised Q-method, 
as shown in Figure \ref{ngc6530_qfit}.
In addition the uncertainties in distance are significantly smaller when 
using the revised Q-method.
The same arguments apply for NGC2244 and $\lambda$ Ori, and 
in Section \ref{results} all three
clusters are fitted using extinctions from the revised Q-method.
For completeness the parameters derived for these three
clusters using both methods are presented in Section
\ref{results_app}. 
The resulting distance moduli derived using a mean extinction or the
revised Q-method are consistent within the uncertainties.

We also attempted to use the revised Q-method for h Per, as it
satisfied our criteria for non-uniform redenning.
However, we found the $E(B-V)$ to vary systematically as a
function of colour. 
This systematic shift in extinction with colour is evident in Figure
\ref{hper_fitebv}.
It shows that to fit the hotter stars and cooler
stars simultaneously would require a change in the gradient of the
isochrone.
The same trend is observed for $\chi$ Per in Figure
\ref{chiper_fitebv}.
Therefore, we attribute this behaviour to differences between the
photometric systems of \cite{2002ApJ...576..880S} and
\cite{1998A&A...333..231B}.
Moreover, as this systematic shift in $E(B-V)$ is not present in
$V$ vs $B-V$ fit (see Figure \ref{chiper_fit}) we further
constrain the problem to a difference dominated by the $U$ band.

\section{Model dependency}
\label{results_app}

Our results for all models are given in Tables
\ref{distances_tycho}-\ref{distances_Q}. The fits have been optimised
by adjusting the systematic uncertainties in colour and magnitude such
that the $Pr(\tau)\approx50\%$ (actually 44\%-66\%).  The values for
the reddening, the distance modulus with 68\% confidence limits and,
the added systematic uncertainties are shown.
Table \ref{distances_tycho} shows the distances derived for $\sigma$
Orionis after conversion to the $TYCHO$ photometric system of
\cite{2000PASP..112..961B}.
Tables \ref{distances_int} and \ref{distances_int_2} contain the
results fitting using a direct conversion from
$T_{eff}$ to colour.
The distances derived for each SFR using the mean extinction
and revised Q-method for each set of isochrones are shown in Tables
\ref{distances_UBV} and \ref{distances_Q}.

\begin{figure}
  \vspace*{174pt}
  \includegraphics{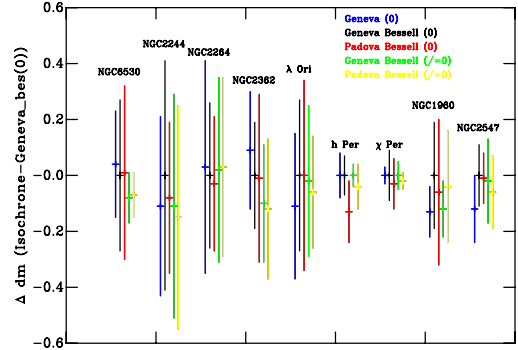}
\caption{The distances and uncertainties (68\% confidence intervals)
  derived for all SFRs and all isochrones, where a mean extinction has
  been derived. The results shown for each cluster are for, from left
  to right, Geneva (0) (blue), Geneva Bessell (0) (black), Padova
  Bessell (0) (red), Geneva Bessell (/=0) (green) and Padova Bessell
  (/=0) (yellow). All the resulting distances agree within the
  uncertainties for all models. \label{diff}}
\end{figure}

These results allow us to perform a brief comparison of the
isochrones we have used, although not the main aim of this paper, it
may aid the reader in adopting a particular result for a SFR of
interest. Figure \ref{diff} shows the derived distances and associated
uncertainties (68\% confidence intervals) for each SFR and each set of
isochrones, where a mean extinction has been derived. 
Figure \ref{diff} shows that for any given SFR the scatter between
models is smaller than the uncertainties from the data.
However, one could argue for a systematic shift of $\approx0.05$ in
distance modulus depending on the choice of Vega zero point.

\begin{table}
\begin{tabular}{lccc}
\hline
\hline
\multicolumn{4}{|c|}{$TYCHO$ photometry}\\
\hline
SFR&\multicolumn{3}{|c|}{$\sigma$ Ori}\\
Model (Vega
cal)&dm&$\Delta$dm&Unc\\
\hline
Geneva-Bessell(0)&$7.84<7.94<8.10$&0.26&0.00\\
Padova-Bessell(0)&$7.72<7.93<8.03$&0.31&0.00\\
Geneva-Bessell($/=0$)&$7.79<7.94<8.07$&0.28&0.005\\
Padova-Bessell($/=0$)&$7.72<7.93<7.98$&0.26&0.005\\
\hline
\hline
\end{tabular}
\caption{The distance moduli derived for $\sigma$ Ori in
  the $TYCHO$ photometric system for the different models, and
  calibrations to Vega colours of zero or Vega $B-V=-0.002$ and
  $U-B=-0.004$. Uncertainties required to achieve
  $0.40<P_r(\tau^2)<0.60$ are also
  shown.\label{distances_tycho}}
\end{table}

\begin{table}
\begin{tabular}{lccc}
\hline
\hline
\multicolumn{4}{|c|}{Intrinsic colours from $T_{eff}$}\\
\hline
SFR&\multicolumn{3}{|c|}{The ONC}\\
Model (colour used)&dm&$\Delta$dm&Unc\\
\hline
Geneva-Bessell($B-V$)&$7.91<7.96<8.03$&0.12&0.010\\
Geneva-Bessell($V-I$)&$7.88<7.96<8.00$&0.12&0.013\\
\hline
\hline
\end{tabular}
\caption{The distance moduli derived for the ONC
  using intrinsic colours derived from $T_eff$ values of
  \citet{1997AJ....113.1733H} in different colour
  indices. Uncertainties required to achieve $0.40<P_r(\tau^2)<0.60$
  are also shown.\label{distances_int}}
\end{table}

\begin{table*}
\begin{tabular}{lcccccc}
\hline
\hline
\multicolumn{7}{|c|}{Intrinsic colours from $T_{eff}$, $B-V$}\\
\hline
SFR&\multicolumn{3}{|c|}{$\chi$ Per}&\multicolumn{3}{|c|}{h Per}\\
Model (stars used)&dm&$\Delta$dm&Unc&dm&$\Delta$dm&Unc\\
\hline
Geneva-Bessell(Hot)&$11.70<11.78<11.93$&0.23&0.003&$11.76<11.78<11.94$&0.18&0.004\\
Geneva-Bessell(Hot and cool)&$11.83<11.84<11.87$&0.04&0.00&$11.83<11.83<11.84$&0.01&0.019\\
\hline
\hline
\end{tabular}
\caption{The distance moduli derived for h and $\chi$
  Per using intrinsic colours derived from $T_eff$ values of
  \citet{2002ApJ...576..880S} alone, or supplemented by cooler stars
  dereddened using an average extinction. Uncertainties required to
  achieve $0.40<P_r(\tau^2)<0.60$ are also shown.\label{distances_int_2}}  
\end{table*}

\begin{landscape}
\begin{table}
\begin{tabular}{lcccccccccccc}
\hline
\hline
\multicolumn{13}{|c|}{Reddening from $UBV$ fitting}\\
\hline
SFR&\multicolumn{4}{|c|}{NGC6530}&\multicolumn{4}{|c|}{NGC2244}&\multicolumn{4}{|c|}{NGC2264}\\
Model (Vega
cal)&dm&$\Delta$dm&Unc&$E_{(B-V)}$&dm&$\Delta$dm&Unc&$E_{(B-V)}$&dm&$\Delta$dm&Unc&$E_{(B-V)}$\\
\hline
Geneva(0)&$10.29<10.38<10.48$&0.19&0.0330&0.35&$10.62<10.78<10.94$&0.32&0.0470&0.47&$9.15<9.40<9.53$&0.38&0.018&0.06\\
Geneva-Bessell(0)&$10.16<10.34<10.43$&0.27&0.0305&0.32&$10.68<10.89<11.09$&0.41&0.0450&0.46&$9.26<9.37<9.52$&0.26&0.0160&0.04\\
Padova-Bessell(0)&$10.15<10.35<10.46$&0.31&0.0308&0.32&$10.73<10.81<11.00$&0.27&0.0450&0.45&$9.27<9.34<9.51$&0.24&0.0140&0.04\\
Geneva-Bessell($/=0$)&$10.25<10.27<10.34$&0.09&0.0330&0.32&$10.54<10.78<10.94$&0.40&0.0400&0.46&$9.18<9.39<9.51$&0.33&0.0160&0.04\\
Padova-Bessell($/=0$)&$10.26<10.28<10.32$&0.08&0.0330&0.32&$10.50<10.74<10.90$&0.40&0.0400&0.45&$9.18<9.40<9.50$&0.32&0.0160&0.04\\       
\hline
\hline 
SFR&\multicolumn{4}{|c|}{NGC2362}&\multicolumn{4}{|c|}{$\lambda$
  Ori}&\multicolumn{4}{|c|}{h Per}\\
Model (Vega
cal)&dm&$\Delta$dm&Unc&$E_{(B-V)}$&dm&$\Delta$dm&Unc&$E_{(B-V)}$&dm&$\Delta$dm&Unc&$E_{(B-V)}$\\
\hline
Geneva(0)&$10.48<10.58<10.69$&0.21&0.0150&0.12&$7.77<7.87<8.03$&0.26&0.022&0.12&$11.77<11.78<11.85$&0.08&0.0250&0.57\\
Geneva-Bessell(0)&$10.51<10.67<10.70$&0.19&0.0130&0.10&$7.89<7.98<8.16$&0.27&0.025&0.11&$11.77<11.78<11.84$&0.07&0.0260&0.54\\
Padova-Bessell(0)&$10.47<10.66<10.77$&0.30&0.0120&0.10&$7.81<7.98<8.15$&0.34&0.026&0.11&$11.63<11.65<11.74$&0.11&0.0310&0.52\\
Geneva-Bessell($/=0$)&$10.49<10.57<10.70$&0.21&0.0120&0.10&$7.86<7.96<8.13$&0.27&0.024&0.11&$11.76<11.78<11.80$&0.04&0.0270&0.54\\ 
Padova-Bessell($/=0$)&$10.47<10.55<10.72$&0.25&0.0120&0.10&$7.79<7.92<7.99$&0.20&0.025&0.11&$11.67<11.74<11.75$&0.08&0.0300&0.52\\
\hline
\hline
SFR&\multicolumn{4}{|c|}{$\chi$ Per}&\multicolumn{4}{|c|}{NGC1960}&\multicolumn{4}{|c|}{NGC2547}\\
Model (Vega
cal)&dm&$\Delta$dm&Unc&$E_{(B-V)}$&dm&$\Delta$dm&Unc&$E_{(B-V)}$&dm&$\Delta$dm&Unc&$E_{(B-V)}$\\
\hline
Geneva(0)&$11.82<11.82<11.85$&0.03&0.010&0.52&$10.21<10.22<10.30$&0.09&0.0170&0.22&$7.85<7.93<7.97$&0.12&0.012&0.053\\
Geneva-Bessell(0)&$11.79<11.82<11.88$&0.09&0.005&0.50&$10.27<10.35<10.46$&0.19&0.0164&0.20&$7.98<8.05<8.09$&0.11&0.018&0.038\\
Padova-Bessell(0)&$11.77<11.79<11.86$&0.09&0.012&0.50&$10.17<10.29<10.43$&0.26&0.0164&0.20&$7.97<8.04<8.06$&0.09&0.018&0.038\\
Geneva-Bessell($/=0$)&$11.80<11.82<11.85$&0.05&0.011&0.50&$10.21<10.23<10.31$&0.10&0.0164&0.20&$7.92<8.03<8.07$&0.15&0.020&0.034\\
Padova-Bessell($/=0$)&$11.79<11.80<11.82$&0.03&0.010&0.50&$10.20<10.31<10.40$&0.20&0.0164&0.20&$7.92<7.99<8.05$&0.13&0.020&0.034\\
\hline
\hline
\end{tabular}
\caption{The distance moduli and $E(B-V)$ values
  , derived using $U-B$ $B-V$ fitting, for each SFR using the
  different models and calibrations to Vega colours of zero or
  Vega $B-V=-0.002$ and $U-B=-0.004$. Uncertainties required to
  achieve $0.40<P_r(\tau^2)<0.60$ are also
  shown.\label{distances_UBV}}
\end{table}

\begin{table}
\begin{tabular}{lcccccccccccc}
\hline
\hline
\multicolumn{13}{|c|}{Revised Q-method}\\
\hline
SFR&\multicolumn{4}{|c|}{NGC6530}&\multicolumn{4}{|c|}{NGC2244}&\multicolumn{4}{|c|}{$\lambda$
  Ori}\\
Model (Vega
cal)&dm&$\Delta$dm&Unc&$\overline{E_{(B-V)}}$&dm&$\Delta$dm&Unc&$\overline{E_{(B-V)}}$&dm&$\Delta$dm&Unc&$\overline{E_{(B-V)}}$\\
\hline
Geneva-Bessell(0)&$10.49<10.50<10.60$&0.11&0.011&0.33&$10.66<10.77<10.81$&0.15&0.01&0.44&$7.99<8.01<8.12$&0.13&0.005&0.10\\
Padova-Bessell(0)&$10.47<10.48<10.58$&0.11&0.011&0.35&$10.68<10.76<10.84$&0.16&0.013&0.45&$7.93<8.01<8.09$&0.16&0.007&0.10\\
Geneva-Bessell($/=0$)&$10.36<10.41<10.49$&0.13&0.011&0.34&$10.67<10.76<10.89$&0.22&0.014&0.42&$7.95<7.97<8.11$&0.16&0.007&0.17\\
Padova-Bessell($/=0$)&$10.48<10.52<10.57$&0.09&0.009&0.34&$10.70<10.81<10.92$&0.22&0.013&0.43&$7.94<7.97<8.08$&0.14&0.007&0.12\\
\hline
\hline
\end{tabular}
\caption{The distance moduli and mean $E(B-V)$ values, after 
application of Q, for each SFR using the
  different models, and calibrations to Vega colours of zero or
  Vega $B-V=-0.002$ and $U-B=-0.004$. Uncertainties required to
  achieve $0.40<P_r(\tau^2)<0.60$ are also
  shown.\label{distances_Q}}
\end{table}
\end{landscape}

These results show that any model dependency in the extinctions and
distances derived is small. A final statistical justification can be
found from the results in Section \ref{results_app}. Here the systematic 
uncertainties required to achieve a $\tau^2$ of
approximately one do not significantly favour any particular model.

\section{Results}
\label{results}

\begin{figure}
  \vspace*{174pt}
  \includegraphics{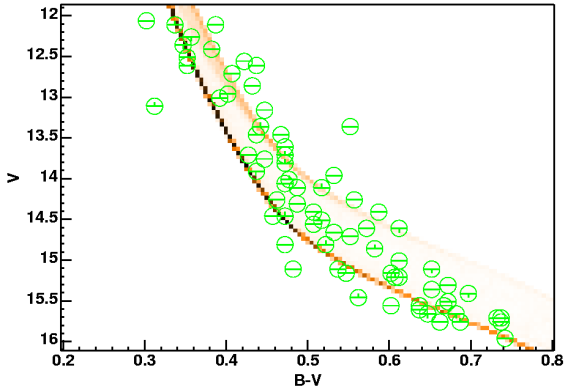}
\caption{h Per distance fit.\label{hper_fit}}
\end{figure}

\begin{figure}
  \vspace*{174pt}
  \includegraphics{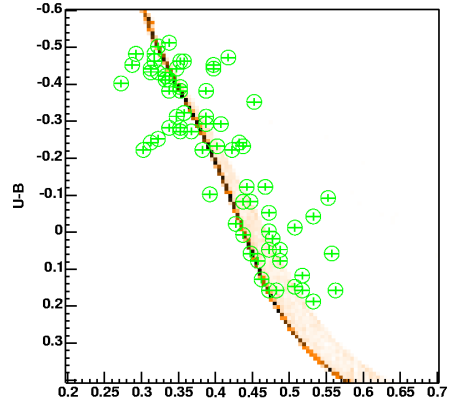}
\caption{h Per $E(B-V)$ fit. Here a systematic shift in $E(B-V)$ as a
  function of colour ($U-B$) is evident, as in Figure
  \ref{chiper_fitebv}. This is probably due to a difference in the
  photometric systems of \citet{1998A&A...333..231B} and
  \citet{2002ApJ...576..880S} (see Section
  \ref{Q}).\label{hper_fitebv}}
\end{figure}

\begin{figure}
  \vspace*{174pt} 
\includegraphics{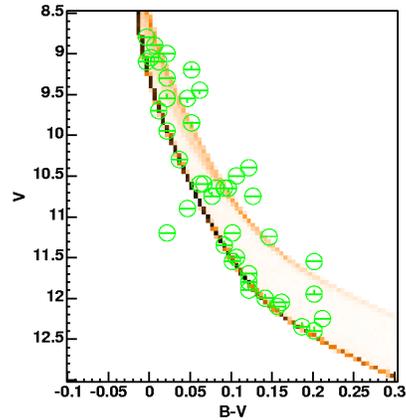}
\caption{NGC1960 distance fit.\label{ngc1960_fit}}
\end{figure}

\begin{figure}
  \vspace*{174pt}
  \includegraphics{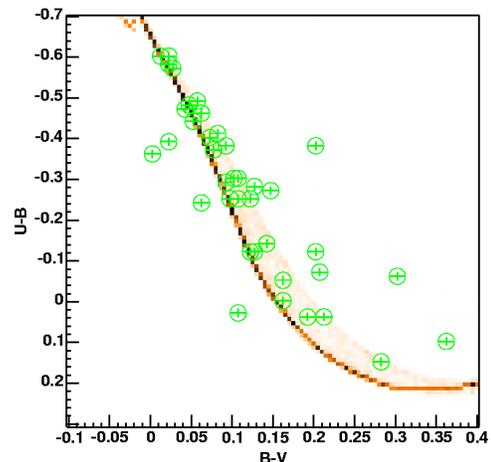}
\caption{NGC1960 $E(B-V)$ fit.\label{ngc1960_fitebv}}
\end{figure}

\begin{figure}
  \vspace*{174pt}
  \includegraphics{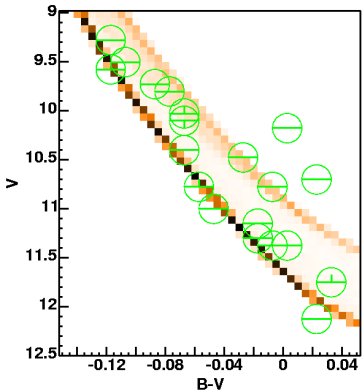}
\caption{NGC2362 distance fit.\label{ngc2362_fit}}
\end{figure}

\begin{figure}
  \vspace*{174pt}
  \includegraphics{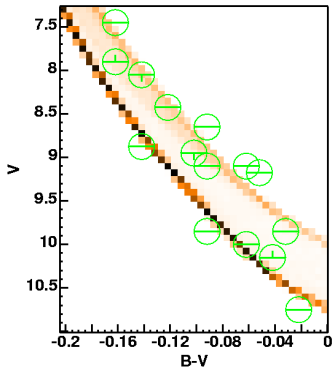}
\caption{NGC2264 distance fit.\label{ngc2264_fit}}
\end{figure}

\begin{figure}
  \vspace*{174pt} 
\includegraphics{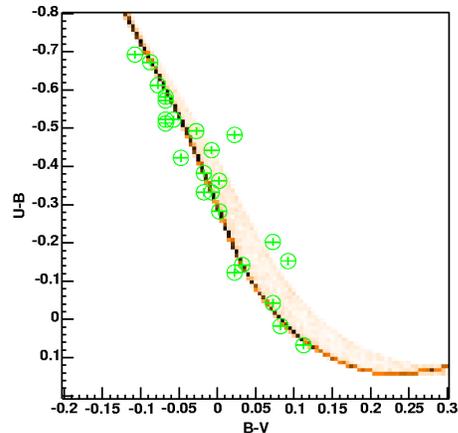}
\caption{NGC2362 $E(B-V)$ fit.\label{ngc2362_fitebv}}
\end{figure}

\begin{figure}
  \vspace*{174pt}
  \includegraphics{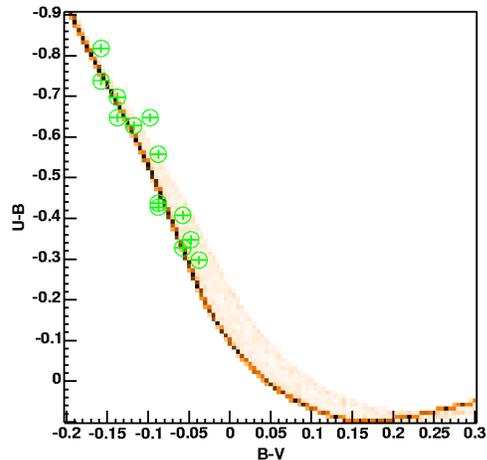}
\caption{NGC2264 $E(B-V)$ fit.\label{ngc2264_fitebv}}
\end{figure}

\begin{figure}
  \vspace*{174pt}
  \includegraphics{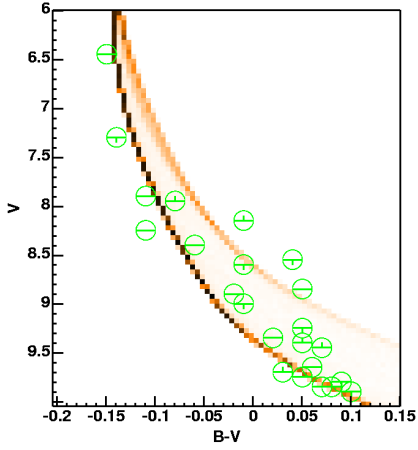}
\caption{NGC2547 distance fit.\label{ngc2547_fit}}
\end{figure}

\begin{figure}
  \vspace*{174pt}
  \includegraphics{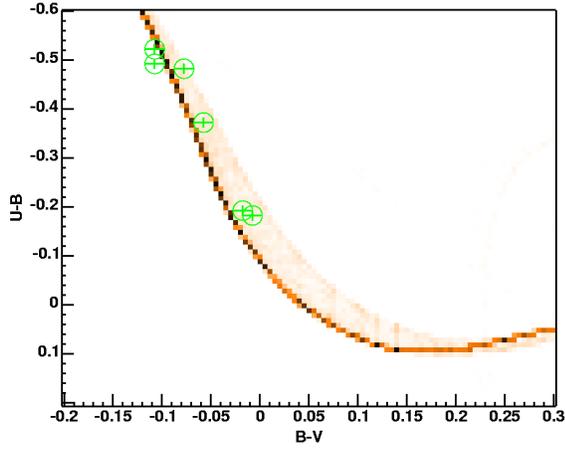}
\caption{NGC2547 $E(B-V)$ fit.\label{ngc2547_fitebv}}
\end{figure}

To simplify our discussion of the implications of our derived distances, we 
have adopted the results from the Geneva-Bessell isochrones. 
We display the resulting fits as Figures \ref{chiper_fit},
\ref{chiper_fitebv} and \ref{hper_fit}-\ref{ngc2547_fitebv}.  In Table
\ref{dist_comp} we provide a comparison of the adopted distances with 68\% 
confidence intervals, with the distances assumed in
\cite{2007MNRAS.375.1220M}. The best fitting $E(B-V)$ is also
provided.

\begin{table*}
\begin{tabular}{lccccccc}
\hline
\multirow{2}{*}{SFR}&\multicolumn{2}{|l|}{\citet{2007MNRAS.375.1220M}}&\multicolumn{2}{|l|}{This
  work}&\multirow{2}{*}{$\Delta\overline{dm}$}&\multirow{2}{*}{$\Delta$ range}&\multirow{2}{*}{$E(B-V)$}\\
&dm&Range&dm&range&\\
\hline
the ONC&$8.01<8.38<8.75$&0.76&$7.91<7.96<8.03$&0.12&-0.42&-0.64&$\approx0.40$$^{(1)}$\\
NGC6530&10.48$^{(4)}$&$\approx0.40$$^{(4)}$&$10.15<10.34<10.44$$^{(2)}$&0.29&-0.14&-0.11&0.32\\
NGC6530&10.48$^{(4)}$&$\approx0.40$$^{(4)}$&$10.49<10.50<10.60$$^{(5)}$&0.11&+0.02&-0.29&0.33$^{(5)}$\\
NGC2244&$10.55<10.72<10.87$$^{(3)}$&0.33&$10.68<10.89<10.94$$^{(2)}$&0.26&+0.17&-0.07&0.46\\
NGC2244&$10.55<10.72<10.87$$^{(3)}$&0.33&$10.66<10.77<10.81$$^{(5)}$&0.15&+0.05&-0.18&0.44$^{(5)}$\\
NGC2264&9.6$^{(6)}$&$\approx0.40$$^{(4)}$&$9.26<9.37<9.52$&0.26&-0.23&-0.14&0.04\\
NGC2362&$10.84<10.87<10.90$&0.06&$10.51<10.67<10.70$&0.19&-0.20&+0.13&0.10\\
$\lambda$ Ori&$7.73<7.90<8.07$&0.34&$7.99<8.01<8.12$$^{(5)}$&0.13&+0.11&-0.21&0.10$^{(5)}$\\
$\lambda$ Ori&$7.73<7.90<8.07$&0.34&$7.89<7.98<8.16$$^{(2)}$&0.27&+0.08&-0.07&0.11\\
$\sigma$
Ori&$7.41<7.80<8.19$&0.78&$7.84<7.94<8.10$&0.26&+0.14&-0.52&0.06$^{(7)}$\\
$\chi$ Per&$11.69<11.70<11.71$&0.20&$11.79<11.82<11.88$&0.09&+0.12&-0.11&0.50\\
h Per&$11.69<11.70<11.71$$^{(8)}$&0.20&$11.77<11.78<11.84$&0.07&+0.08&-0.13&0.54\\
NGC1960&$10.40<10.60<10.80$$^{(9)}$&0.40&$10.27<10.35<10.46$&0.19&-0.25&-0.21&0.20\\
NGC2547&$7.92<8.18<8.47$$^{(10)}$&0.54&$7.98<8.05<8.09$&0.11&-0.13&-0.43&0.038\\
\hline
\end{tabular}
\caption{Distance moduli and range assumed
  in \citet{2007MNRAS.375.1220M} and those derived in this work. 
  Notes are as follows.
  (1) Individual extinctions from $T_{eff}$, value given is approximate
  mean.
  (2) Distance derived using mean extinction derivation from
  $\tau^2$ fitting. 
  (3) Not included in \citet{2007MNRAS.375.1220M}, distance from 
  \citet{2000A&A...358..553H}. 
  (4) Typical uncertainties assumed as none were provided in
  literature source.
  (5) Resulting distance after application of extinctions derived using
  Q-method (see Section \ref{Q}). A mean extinction is quoted.
  (6) Typical uncertainties assumed as literature distance is the mean
  of many values.
  (7) $E(B-V)$ from \citet{1994A&A...289..101B}.
  (8) Assumed to be at the same distance as $\chi$ Per in \citet{2007MNRAS.375.1220M}.
  (9) Not included in \citet{2007MNRAS.375.1220M}, distance from 
  \citet{2002AJ....123..892P}.
  (10) Not included in \citet{2007MNRAS.375.1220M}, distance from
  \citet{1999A&A...345..471R}. \label{dist_comp}} 
\end{table*}

\subsection{Notes on results}
\label{odd_res}

We would have liked to include the sub-group CepOB3b in this paper,
but after application of the Q-method the resulting colour range for
the stars available for fitting was prohibitively low for distance
fitting. Literature derivations of extinction 
\citep{1970AJ.....75.1001G,1959ApJ...130...69B} rely on intrinsic
colours derived from other isochrones so cannot be used to fit with
the Geneva-Bessell isochrones.

\begin{figure}
  \vspace*{174pt} 
\includegraphics{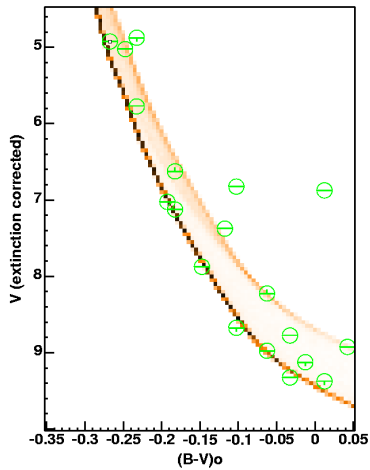}
\caption{The ONC distance fit. Figure shown in intrinsic colour and
  extinction corrected magnitude (see Section
  \ref{ind_ext}).\label{onc_fit}}
\end{figure}

\begin{figure}
  \vspace*{174pt}
  \includegraphics{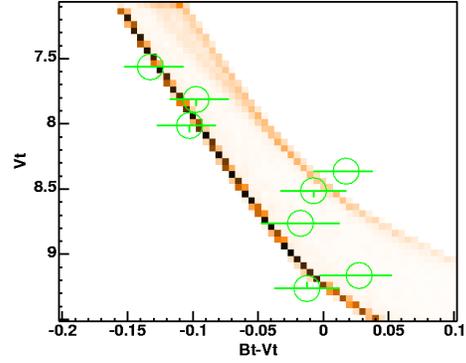}
\caption{$\sigma$ Orionis distance fit. \label{sori_fit}}
\end{figure}

\begin{figure}
  \vspace*{174pt} 
\includegraphics{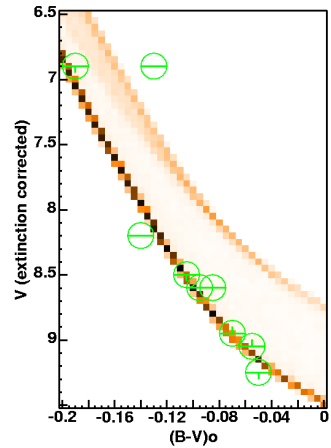}
\caption{$\lambda$ Ori distance fit using the Q-method extinctions
  (see Section \ref{Q}). Figure shown in intrinsic colour and extinction corrected
  magnitude (see Section \ref{ind_ext}).\label{lambdaori_qfit}}
\end{figure}

\begin{figure}
  \vspace*{174pt}
  \includegraphics{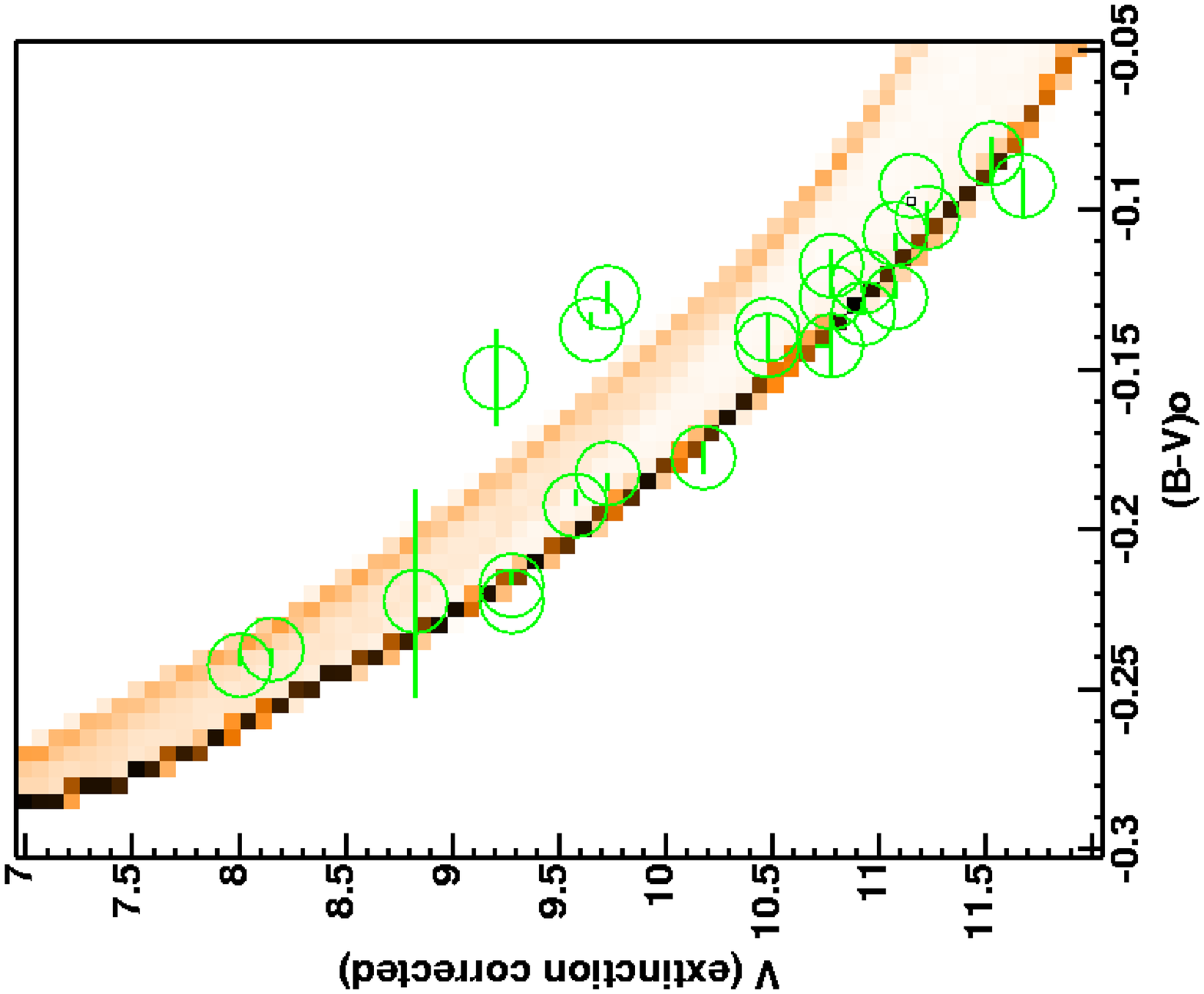}
\caption{NGC2244 distance fit using Q-method extinctions (see Section
  \ref{Q}). Figure shown in intrinsic colour and extinction corrected
  magnitude (see Section \ref{ind_ext}).\label{ngc2244_qfit}}
\end{figure}

\section{Implications}
\label{implications}

We have now derived a self-consistent set of distances (and
extinctions) to, in general, a higher precision than that existing in
the literature, with statistically meaningful uncertainties for these
distances. 
We now discuss some of the key
implications of both the individual distances and of the
entire dataset.

\subsection{Individual Distances}
\label{imp_ind}

Of the SFRs studied in this work distances derivations for eight are
of particular note. Here we have converted the distance moduli to a
distance to allow more obvious comparisons.

\subsubsection{The ONC}
\label{imp_onc}
   
We have increased the precision of the distance estimate for the ONC
by a factor of 7 compared to that used in \cite{2007MNRAS.375.1220M}.
This new distance is also closer than the previously accepted result
from the maser measurements of \cite{1981ApJ...244..884G}, by $0.42$
mag. Conversely this new distance, $391^{+12}_{-9}$ pc agrees superbly
with several recent derivations in the literature. Firstly,
\cite{2007MNRAS.376.1109J} finds a distance of $392\pm32$ pc from the
rotational properties of low-mass pre-MS stars (after removing
accreting objects).  Secondly, a parallactic distance of
$389^{+24}_{-21}$ pc from very long baseline array observations has
been found by \cite{2007arXiv0706.2361S}.  Lastly,
\cite{2007A&A...466..649K} find a distance of $434\pm12$ or $387\pm11$
pc by modeling the orbit of the $\theta^1$Ori C binary system. They
adopt 434 pc as the likely result after comparison to the distance
obtained by \cite{2007MNRAS.376.1109J} for all objects including those
showing evidence of accretion. Clearly, our result favours the
solution yielding 387 pc. A convenient round number which agrees with
the majority of the recent derivations is 400 pc.

\begin{figure}
  \vspace*{174pt}
  \includegraphics{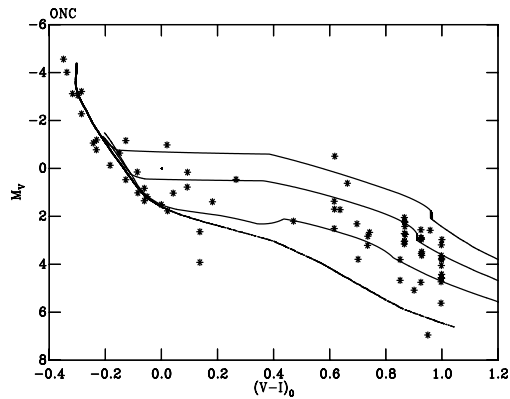}
\caption{The ONC with a Geneva-Bessell 1 Myr MS isochrone and the Pre-MS
  isochrones of \citet{2000A&A...358..593S} for 1, 3 and 10 Myrs.
  Stars appear to lie at the turn-on for an age of 10
  Myr.\label{spread_young}}
\end{figure}

This closer distance ($391^{+12}_{-9}$ pc compared to $480\pm80$ pc)
has important implications for the stellar population of the ONC. It
means the pre-MS population lies $0.42$ mags fainter in absolute
magnitude in the CMD.  This will force the isochronal age of stars
older, but perhaps more importantly increase their spread 
in isochronal age derived from a CMD
\citep[see the discussion in ][]{2005ApJ...626L..49P}. 
This is due to the bunching of older isochrones towards
the zero-age-main-sequence (ZAMS). In fact there is also evidence of a
spread in the CMD from the MS members we have used to derive a
distance.  As can be seen in Figure \ref{spread_young}, MS stars exist
at a position in the CMD suggesting isochronal ages of up to 10 Myrs,
which is at variance with the median pre-MS age of $\approx2$ Myr
(after allowing for the revised distance).
This is discussed further in Section \ref{imp_cuts}.

\subsubsection{$\sigma$ Orionis}
\label{imp_sori}

We have improved the precision of the distance estimate used for
$\sigma$ Orionis in \cite{2007MNRAS.375.1220M} by a factor of 2.
\cite{Caballero.2007.} derives a distance based on $TYCHO$ photometry
of $360^{+70}_{-60}$ pc. Although we also use the $TYCHO$ photometry,
we use updated photometric conversions from
\cite{2000PASP..112..961B}. We derive a distance of $389^{+34}_{-24}$
pc, a value more precise than, but in agreement with that of
\cite{Caballero.2007.}.

\subsubsection{NGC2547}
\label{imp_ngc2547}

For NGC2547 we have increased the precision in distance from that
adopted in \cite{2007MNRAS.375.1220M} by a factor of 5. In this work
we derive a distance of $407^{+8}_{-13}$ pc, which compares favourably
with the $HIPPARCOS$ result of $433^{+62}_{-49}$ pc from
\cite{1999A&A...345..471R}. However, \cite{2006MNRAS.373.1251N} use
pre-MS isochrones to obtain a distance of $361^{+19}_{-8}$ pc. The age
and distance derivation in \cite{2006MNRAS.373.1251N} is consistent
with the Li depletion boundary, which is also based on pre-MS models.
Therefore we conclude the difference in distance is attributable to a
model-dependent difference between the MS and pre-MS models.
Interestingly we also obtain a different reddening, $E(B-V)=0.038$ to
that of \cite{Claria.1982.}, $E(B-V)=0.06\pm0.02$ which is used in
\cite{2006MNRAS.373.1251N}. However, this does not have a significant
impact on their distance, and so does not explain the discrepancy
between the MS and pre-MS distances.

\subsubsection{NGC2244}
\label{imp_ngc2244}

The distance to NGC2244 derived in this work is $1\,425^{+27}_{-70}$
pc, using individual extinctions from the revised Q-method (see Section
\ref{Q}).  This compares well to the literature result of
$1\,390\pm100$ pc from eclipsing binaries of
\cite{2000A&A...358..553H}. Previous MS isochrone fitting studies
placed this SFR at $1\,667^{+128}_{-118}$ pc
\citep{1987PASP...99.1050P}, and $1\,660$ pc
\citep[][we use the same data as this
study]{2002AJ....123..892P}. Our result confirms the closer distance of
\cite{2000A&A...358..553H} and is marginally consistent with the
studies yielding a greater distance.
The new distance will move the cluster pre-MS fainter or older by 
$\approx0.3$ mags.

\subsubsection{NGC2362}
\label{imp_ngc2362}

Our derived distance for NGC2362, $1\,361^{+19}_{-97}$ pc is closer
and less precise than that used in \cite{2007MNRAS.375.1220M}.  In
\cite{2007MNRAS.375.1220M} we adopted a distance of
$1\,493^{+21}_{-20}$ pc, an incredibly precise value from
\cite{1996MNRAS.281.1341B}, derived using a form of MS fitting to
narrow band photometry. For the sake of a consistent relative
experiment we adopt our new distance for subsequent analysis.

\subsubsection{h and $\chi$ Per}
\label{imp_per}

The distance we have derived for $\chi$ Per of $2\,312^{+65}_{-32}$ pc
agrees well with the recent literature result of $2\,344^{+55}_{-53}$
pc of \cite{2002ApJ...576..880S}, as stated in Section
\ref{chiper_eg}.  They employ spectroscopy as well as the Q-method
(with what appears to be an updated MS line, but canonical reddening
vectors) combined with MS isochrone fitting to derive this distance.
Interestingly they also derive a spectroscopic parallax distance of
$3\,162$ pc. This method involves dereddening stars with known
spectral types onto an intrinsic MS; they conclude that recalibration
of these intrinsic colours is required. Also
\cite{2002ApJ...576..880S} find h and $\chi$ Per to be at
approximately the same distance, agreeing with the majority of the
literature \citep[e.g][]{2001AJ....122..248K}. Our distance derivation
for h and $\chi$ Per are also consistent with both these clusters
being at the same distance.

\subsection{Global issues}
\label{imp_glob}

In this section we discuss general implications related to the dataset
as a whole.

\subsubsection{Metallicity} 
\label{imp_met}

There is little work on the metallicity of SFRs.
\cite{2006A&A...446..971J} show that the metallicity is solar or very
slightly sub-solar for the majority of stars in the Lupus, Chamaeleon 
and CrA SFRs.
Conversely recent eclipsing binary results suggest
approximately half-solar metallicity for h and $\chi$ Per and Collinder 228
\citep{2004MNRAS.355..986S, 2004MNRAS.349..547S, 2007A&A...461.1077S},
and solar metallicity for NGC6871
\citep[$Z=0.02$]{2004MNRAS.351.1277S}. If compositions do indeed vary
as is suggested from the above results then distances derived to these
SFRs must be re-derived after a comprehensive composition survey. As
discussed in Section \ref{composition}, adopting a half-solar
composition for $\chi$ Per (for example) results in a fall of the
derived distance modulus by $\approx0.5$ mags. A similar result would
apply for h Per. 
This would have a severe impact on any age ladder, and thus on
any conclusions about secular evolution, such as disc lifetimes.

\subsubsection{Age spreads and the R-C gap overlap}
\label{imp_cuts}

\begin{figure}
  \vspace*{174pt}
  \includegraphics{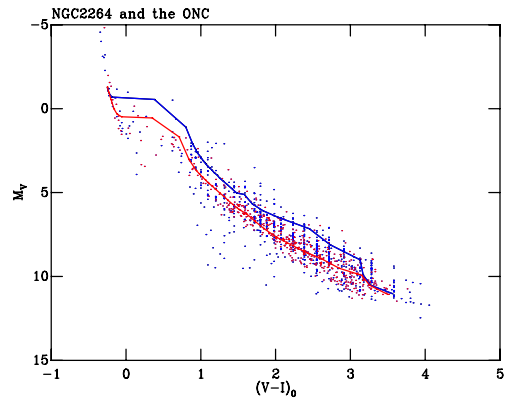}
\caption{Absolute magnitude as a function of intrinsic colour 
for stars in the ONC and NGC2264 \citep[taken from][and
  sources referenced therein]{2007MNRAS.375.1220M} adjusted using the
  extinctions and distances from this work. The blue dots are the ONC
  and the red dots are NGC2264. The locus of points for NGC2264 can be
  seen to lie below the ONC, however the MS section is very similar in
  extent for each SFR, i.e. the turn-on appears coincident. The pre-MS
  isochrones from \citet{2000A&A...358..593S} are shown for 1 and 3
  Myrs. Most importantly in the MS section stars from both populations
  extend below the turn-ons whilst for both SFRs the peak of the MS
  (brightest stars) is similar i.e. the turn-offs do not appear
  significantly different, although this is based on only a few
  stars.\label{onc_ngc2264}}
\end{figure}

As can be seen in Table \ref{prep} the faint red edge of the MS
defined using pre-MS isochrones and the edge of the apparent MS in the
CMD do not agree for the younger SFRs.
This means that there are MS stars which lie fainter and redder than the
base of the MS predicted by theory.
The models predict that, for a coeval population, the brightest stars
still on the pre-MS (at the red edge of the R-C gap) are similar in
magnitude to the faintest stars on the MS (which are at the blue edge
of the R-C gap).
This extension of the MS to magnitudes fainter than the head of the
pre-MS we term the R-C gap overlap.
The R-C gap overlap is most apparent in the younger SFRs.
One of the best examples is shown in Figure 
\ref{spread_young}; the faintest stars on the MS in the ONC must be 
at least 10 Myrs old to have reached the MS,
whilst the median age of the pre-MS is around 2 Myrs.  
An apparent age spread can also be seen for NGC2264 in Figure
\ref{onc_ngc2264} for both the MS and pre-MS populations. 

This apparent extension (along a MS line) below the turn-on for almost all
SFRs can be interpreted as an age spread within the SFR. 
This suggests that stars have evolved across the R-C gap before theory
would predict for a coeval population and must therefore be older.
This would explain why the R-C gap overlap is less apparent in the older SFRs.
As the age of an isochrone increases they become fainter and
move towards the ZAMS.  
Lower mass stars on the pre-MS contract more slowly as they age.
Therefore, the same difference in age for an older population produces
a smaller change in $V$ than for a younger population. 
Thus the R-C gap overlap for older SFRs is harder to detect.  

\begin{figure}
  \vspace*{174pt}
  \includegraphics{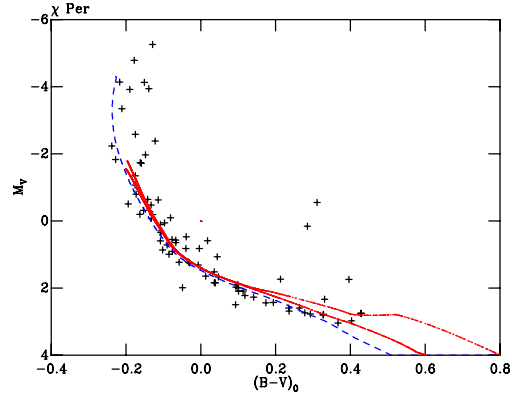}
\caption{$\chi$ Per with a Geneva-Bessell 13 Myr MS isochrone (dashed
  line) and the pre-MS isochrones of \citet{2000A&A...358..593S} for
  13 and 23 Myrs (bold lines). Stars appear to lie below the turn-on
  and above the turn-off for an age of 13 Myr .\label{spread_old}}
\end{figure}

Despite the R-C gap overlap being harder to detect in old SFRs, an 
isochronal age spread can still be seen in the HM stars.  
As shown for $\chi$ Per in Figure \ref{spread_old} stars again lie
below the turn-on and above the apparent turn-off for the best fitting
or literature age of 13 Myrs.  
If one assumes the age is incorrect and increases it to match the
turn-on, the turn-off will move fainter and exacerbate the problem for
the brighter stars.  
Photometric variability and errors, and binarity have been shown not
to completely account for these isochronal age spreads for the pre-MS in
\cite{2005MNRAS.363.1389B} and cannot account for the R-C gap overlap.  

Hypothesising such an age spread supports other results from isochrone
modeling. 
It is well known that individual ages derived using pre-MS isochrone
fitting also show an age spread 
\citep[see for example][]{2005ApJ...626L..49P,2004ApJ...610.1045S}.
Also, \cite{2007arXiv0707.4641J} finds a direct spread in the radii
(and hence by implication age)  of the PMS stars in the ONC, at a given 
effective temperature, a method free from isochrone theory.

However it is dangerous to interpret the R-C gap overlap, or the spreads 
in stellar radii and isochronal age, as real age spreads.
As shown in \cite{1999MNRAS.310..360T} accretion can act to
force a star bluer and temporally older. 
This shift in position within a CMD means an isochronal age spread
derived from pre-MS fitting does not necessarily imply a real
underlying age spread of the same magnitude.
Additionally \cite{1999A&A...342..480S} find that the evolution of an
accreting star is accelerated, with the star having a smaller radius
and therefore lower luminosity than a non-accreting coeval
counterpart. 
This result shows that age spreads derived from spreads in radii or
from the R-C gap overlap again do not necessarily imply a real spread in
age.

Whether star formation is rapid ($\approx1$ Myr) or slow ($\approx
5-10$ Myrs) is currently an active debate
\citep[e.g][]{2007ApJ...654..304K,2007RMxAA..43..123B}.  
Not withstanding different star formation rates or several
episodes of star formation, the age spread of the bulk of
the population within an SFR  is an approximate measure for the local star 
formation time.
Apparent age spreads within a CMD are often used to support the model
of slow star formation
\citep{2005ApJ...626L..49P,2005MNRAS.363.1389B,2007arXiv0707.4641J}.
In the rapid star formation model these spreads of apparent age are dominated by
accretion effects, and the residual real age spread is small.
If accretion does indeed act to scatter a star within the CMD and even
artificially accelerate its evolution or contraction, isochronal ages
should not be used to represent real age spreads.
Moreover, isochronal ages (based on turn-ons, turn-offs or pre-MS
fitting) for individual stars without a known accretion history do not
represent the true age of the star and therefore should not be used to
support evolutionary theories.
Indeed, it is even hard to argue that a median or mean age for a given
SFR derived from isochrone fitting has any real meaning.
Perhaps, following the results in \cite{1999A&A...342..480S} and
\cite{1999MNRAS.310..360T}, if accretion causes a decrease in the
star's radius, therefore increasing its isochronal age, the most
accurate representative age for a given cluster is that of the
youngest stars having the lowest accretion histories.  
However, this would mean a dramatic change in the ages for most SFRs,
for example the youngest stars in the ONC are $\approx0.1$ Myrs
\citep{1997AJ....113.1733H} and many older SFRs still contain active
star formation and embedded objects.  

A useful indicator to quantify these perceived spreads may be the R-C
gap overlap, where the minimum mass object to have developed a
radiative core and to have joined the MS can be compared to the
maximum mass stars still existing on the convective pre-MS.  
This overlap then provides a precise diagnostic for the apparent age
spreads. 
Whether these spreads show a real age spread or are indicative of the
range of accretion histories present depends on the model adopted.
  
Given the problems with SFR ages the best approach currently is
to compare observations of two different SFRs. 
Either deriving an age order by assuming a similar range of accretion
rates within each SFR, or by using the R-C gap overlap to derive
approximate differences in the range of accretion rates.
An example of this can be seen in Figure \ref{onc_ngc2264}, showing
the absolute magnitude and intrinsic colour for the stars of the ONC and 
NGC2264 \citep[taken from][and sources referenced
therein]{2007MNRAS.375.1220M}.
The locus of the pre-MS in NGC2264 clearly lies slightly below that of
the ONC, but the MS section is strikingly similar. 
In addition Figure \ref{onc_ngc2264} shows the MS for the SFRs
extending below the predicted turn-on.  
Moreover the brightest MS stars in both populations are at similar
magnitudes i.e.  the turn-off is in a similar position with stars
lying on the apparent MS above the turn-off (as seen in Figure
\ref{spread_old}). 
As the apparent MS and pre-MS in both clusters appear to extend to
similar points, i.e.  the R-C gap overlap is similar, suggesting a similar
range of accretion histories or ages for each cluster.  
So, in conclusion, the MS and pre-MS sections imply a large
\textbf{isochronal} age spread as seen in Figure \ref{spread_young},
and deriving ages and age spreads from these sections of the sequence
would lead to a similar result for the ONC and NGC2264.
However, comparing the sequences as a whole show that NGC2264 is more
evolved than the ONC.

\subsubsection{Secular evolution and ages}
\label{imp_sec}

We now reconstruct the age ladder of \cite{2007MNRAS.375.1220M} using
the new distances and extinctions from Table \ref{dist_comp}. 
We have however changed the process slightly. 
Given that the new distances are generally more precise, we assume that
the MS for each SFR will be approximately coincident in the CMD.
Therefore, we plot the individual photometric points for this section
of each sequence.
Then for stars redward of the R-C gap we fit a spline through the median
points. 
We retain the ZAMS subtracted space as a presentation tool
\citep[where the colour at each magnitude has the corresponding ZAMS
colour subtracted from it as in][]{2007MNRAS.375.1220M}.
We are unable to include NGC2244 as the pre-MS of this SFR is not
studied in  \cite{2007MNRAS.375.1220M}.
We do however, include in this section four SFRs from
\cite{2007MNRAS.375.1220M} not studied in this work, where we adopt
literature distances and extinctions.

As the distance moduli for some SFRs have changed from those adopted in
\cite{2007MNRAS.375.1220M} the nominal ages may also have changed. 
The distance moduli which have changed by more than $0.2$ mag which were
discussed in \cite{2007MNRAS.375.1220M} are the ONC, NGC2264, NGC2362,
$\lambda$ Ori and $\sigma$ Ori. 
In addition the relative distances and extinctions for h and $\chi$
Per have changed slightly, therefore we have refitted these clusters.
Following \cite{2007MNRAS.375.1220M} we have adjusted h Per to the
distance and extinction of $\chi$ Per and combined the sequences prior
to fitting them. 
We have included the entire sequence in a fit to h and $\chi$ Per
providing an empirical ZAMS blue-ward of the R-C gap.

\begin{figure}
  \vspace*{174pt}
  \includegraphics{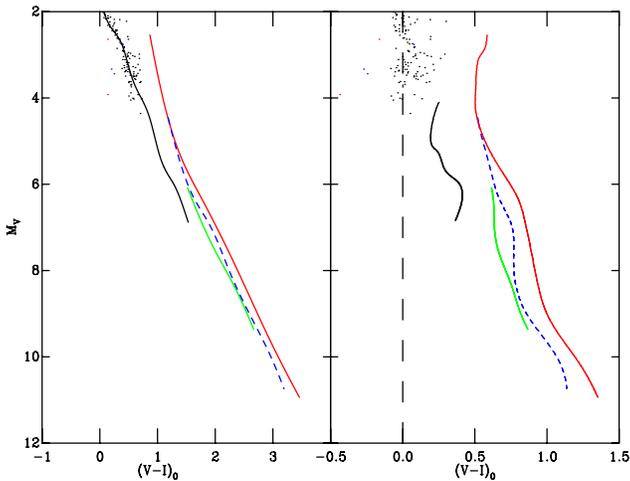}
\caption{The MS stars (dots) and pre-MS empirical isochrones
  (continuous lines) of the ONC (red), h and $\chi$ Per (black) and
  NGC2362 (green), with the pre-MS empirical isochrone of NGC2264
  (blue) as a dashed line. The left panel is a CMD of absolute magnitude and
  intrinsic colour with the
  right panel is in ZAMS subtracted space.\label{rel_one}}
\end{figure}

\begin{figure}
  \vspace*{174pt}
  \includegraphics{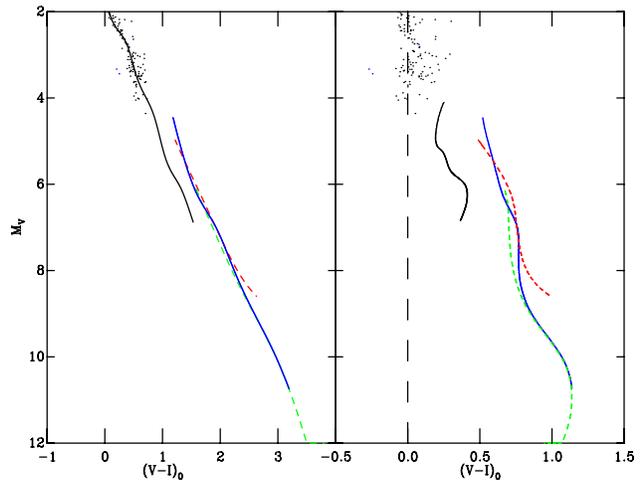}
\caption{The MS stars (dots) and pre-MS empirical isochrones
  (continuous lines) of NGC2264 (blue) and h and $\chi$ Per (black),
  with the pre-MS empirical isochrones of $\lambda$ Ori (red) and
  $\sigma$ Ori (green) as dashed lines. The right panel shows the same
  in ZAMS subtracted space.  The left panel is a CMD of absolute magnitude and
  intrinsic colour with 
  the right panel is in ZAMS subtracted space.\label{rel_two}}
\end{figure}

Following \cite{2007MNRAS.375.1220M} we present fiducial empirical
isochrones bounding the target empirical isochrone in a CMD of 
absolute magnitude and intrinsic colour and in ZAMS subtracted space. 
Figure \ref{rel_one} shows the pre-MS of NGC2264, after application of
the distances in this work, lying only slightly below the ONC, the
pre-MS of NGC2362 is shown to lie below these sequences.  
A combined h and $\chi$ Per empirical isochrone is also shown as a
lower fiducial in Figures \ref{rel_one} and \ref{rel_two}, with Figure
\ref{rel_two} using NGC2264 as an upper fiducial. 
Figure \ref{rel_two} also shows the empirical isochrones for the
pre-MS of both $\lambda$ Ori and $\sigma$ Ori, the former lying above
NGC2264 and the latter below.  
From these two figures we can create an age ladder (youngest to
oldest) and assign nominal ages: the ONC (2 Myrs), $\lambda$ Ori ,
NGC2264 and $\sigma$ Orionis (3 Myrs), NGC2362 (4-5 Myrs) and finally
h and $\chi$ Per (13 Myrs).

We have repeated this method to obtain positions in an age ladder for
each SFR for which we have enough data to create an empirical
isochrone.
This has resulted in the creation of a new group in addition to those
of \cite{2007MNRAS.375.1220M}.
This group, at a nominal age of 2 Myrs, contains the ONC and NGC6530,
as it lies older than IC5146 but marginally younger than NGC2264. 
The resulting age groups from this work and those from
\cite{2007MNRAS.375.1220M}, alongside literature estimates of the
fraction of stars exhibiting infrared (IR) excess (i.e. candidates for
an associated disc), are shown in Table \ref{ladder}. 
It is important to note that we have only updated distances for some
of the SFRs.

\begin{table*}
\begin{tabular}{|l|c|c|l|}
\multirow{2}{*}{SFR}&\multicolumn{2}{|c|}{Nominal
  age}&\multirow{2}{*}{Fraction of stars with IR excess}\\
&This work&\citet{2007MNRAS.375.1220M}&\\
\hline
IC5146&-&1&-\\
NGC6530&2&1&$44\%$$^{(7)}$\\
the ONC&2&1&$80\pm5\%$$^{(1)}$\\
$\lambda$ Ori&3&3&$\approx25\%$,$\approx14\%$,$\approx31\%$$^{(6)}$\\
CepOB3b&-&3&-\\
NGC2264&3&3&$52\pm10\%$$^{(1)}$\\
$\sigma$ Ori&3&4-5&$31.1\pm3.8\%$, $26.6\pm2.8\%$ and $33.9\pm3.1\%$$^{(2)}$\\
NGC2362&4-5&3&$12\pm4\%$$^{(1)}$,$7\pm2\%$$^{(3)}$\\
IC348&-&4-5&$65\pm8\%$$^{(1)}$\\
NGC7160&-&10&$\approx20\%$$^{(4)}$\\
h and $\chi$ Per&13&13&$2-3\%$$^{(8)}$\\
NGC1960&20&20&$3\pm3\%$$^{(1)}$\\
NGC2547&40&38&$\approx7\%$$^{(5)}$\\
\hline
\end{tabular}
\caption{The new nominal ages of SFRs from this work and those from
  \citet{2007MNRAS.375.1220M} are shown in addition to the fraction of
  stars with IR excesses. NGC2244 from this work was not studied in
  \citet{2007MNRAS.375.1220M} so is omitted. IC5146, CepOB3b, IC348 and NGC7160
  are studied in \citet{2007MNRAS.375.1220M} and included here by
  adopting literature distances and extinctions.
  The notes are as follows.
  (1) \citet{2001ApJ...553L.153H}.
  (2) \citet{2007ApJ...662.1067H}. First value for TTS stars
  (approximate mass range $1-0.1M_{\odot}$). Second and third values
  for entire sample, first for thick discs and second for thick and
  evolved discs. 
  (3) \citet{2007AJ....133.2072D}. 
  (4) \citet{2005AJ....130..188S}. 
  (5) \citet{2004ApJS..154..428Y}. 
  (6) \citet{2007ApJ...664..481B}. Firstl value derived from the IRAC
  CCD. Second and third from spectral-energy distribution
 (SED) fitting, thick discs then thin and thick discs combined. All
  values in the approximate mass range $0.1-1.0 M_{\odot}$.
  (7) \citet{2007A&A...462..123P}. 
  (8) \citet{2007ApJ...659..599C}, for h and $\chi$ Per stars fainter
  than J=13.5, stars brighter than this have $0\%$ disc fraction.
  It is important to note that each of these studies is over a
  different mass range (due to different distances and apparent magnitude
  ranges) and excess candidates were selected in a heterogeneous
  fashion.
  \label{ladder}}
\end{table*}

\begin{figure}
  \vspace*{174pt}
  \includegraphics{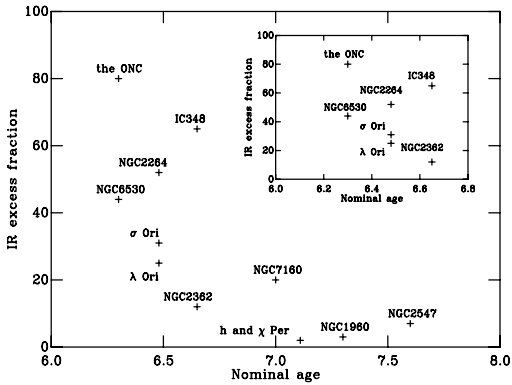}
\caption{Figure showing the log(nominal age) of each SFR with the IR excess
  from various literature sources. The inset panel shows the region
  0-5 Myrs enlarged. \label{disc_order_log}}
\end{figure}

The infrared excess fractions from Table \ref{ladder} can be used to
infer the presence of a disc, then by comparing the disc fraction
across a range of SFRs one can examine the evolution of these discs,
as in \cite{2001ApJ...553L.153H}. Figure \ref{disc_order_log} shows
the logarithmic nominal ages for the SFRs from Table \ref{ladder} and
the inferred fraction of stars with discs.   However the disc fractions inferred
from the data in Table \ref{ladder} come from different
mass ranges, dependent on the apparent magnitude range (and therefore distance)
of the target SFR. Additionally, the excess criteria used are
different, with studies such as \cite{2007ApJ...662.1067H} using the
Spitzer Space Telescope IRAC and MIPS camera channels, whereas disc
fractions in \cite{2001ApJ...553L.153H} were calculated using $JHKL$
excesses.
Therefore, given the heterogeneous nature of these data one cannot draw
strong conclusions from Figure \ref{disc_order_log} regarding the disc
fraction as a function of age. 

Despite its limitations, Figure \ref{disc_order_log} 
is not consistent with a uniform decay, revealing further possible evidence
of environmental effects as suggested in \cite{2007MNRAS.375.1220M}.
As an example we examine the inferred disc fractions of three SFRs in
the same age group (nominal age of 3 Myrs); NGC2264, $\lambda$ Orionis
and $\sigma$ Orionis, with distances of $dm=9.37$, $8.01$ and $7.94$
respectively.  The disc fractions adopted are (in the same order)
$52\pm10\%$ \citep[JHKL,][for masses greater than $0.85
M_{\odot}$]{2001ApJ...553L.153H}, $\approx25\%$ \citep[from IRAC
data,][all discs in spectral range of M0-M6.5 or approximate mass
range of $0.1-0.8 M_{\sun}$ using pre-MS
isochrones]{2007ApJ...664..481B}, and $31.1\pm3.8\%$ \citep[from IRAC
data,][all stars in the approximate mass range $0.1-1.0
M_{\sun}$]{2007ApJ...662.1067H}. In the case of these three SFRs
NGC2264 has an inconsistent disc fraction, it is much higher than that
of the other two SFRs. These disc fractions are taken from differing
mass ranges and the SFRs are at different distances which could lead
to sensitivity problems in the $L$ band as suggested for NGC2362 by 
\cite{2003MNRAS.338..616L}.  However, in this
case the further distance to NGC2264 would result in fewer L band
detections and the lower mass limit being higher is also likely to decrease
the detected disc fraction. Therefore, it is likely that a consistent
experiment would increase the discrepancy between NGC2264 and the two
SFRs with lower disc fractions. However, even ignoring the particular
case at a nominal age of 3 Myrs it is clear that the these data do not
necessarily imply a smooth decline in disc fraction with age,
suggesting other, presumably environmental factors may affect disc
lifetimes.

\section{Conclusions}
\label{imp_sum}

\begin{enumerate}
\item We have derived a self-consistent set of distances of generally
  higher precision than previously available for a set of SFRs in the
  age range 1-40 Myrs. We have also derived distances using several
  other models and  calibrations (see Section \ref{results_app}).
\item In addition to these new distances and reddenings (or
  extinctions) we have reconstructed the age ladder of
  \cite{2007MNRAS.375.1220M} assigning new nominal ages, as shown in
  Table \ref{ladder}. To enable the reader to add other SFRs to this
  ladder the pre-MS splines are freely available from the cluster
  collaboration home
  page\footnote{http://www.astro.ex.ac.uk/people/timn/Catalogues/description.html},
  and the CDS archive.
\item We have shown that metallicity information is now vital for
  accurate relative distances to SFRs. This is especially true if one
  is attempting to characterise evolutionary indicators such as disc
  fractions as a function of age, or trying to uncover environmental
  effects (such as the effect of ionising winds from massive stars on
  planet formation from discs).
\item We have discussed that the overlapping region of the R-C gap
  could be used to derive spreads in isochronal age in a CMD,  i.e. the
  apparent age difference between the 
  maximum mass star still on the convective pre-MS and the minimum
  mass star which has reached the MS. If star formation is slow and
  isochronal ages of individual stars are reliable this would provide
  a direct measurement of the age spreads present in SFRs. If star
  formation is rapid the R-C gap overlap region reveals the underlying
  spread in accretion histories within an SFR. This is important as
  for rapid star formation, if an accretion history is unknown
  isochronal ages derived from a position in a CMD do not represent
  the true age of a star. Indeed it is therefore likely that if a
  rapid star formation model is accurate median or mean ages drawn
  from a population are also invalid. A more useful approach may be to
  compare SFRs using age ladder arguments or perhaps to use the age
  of the youngest stars which have the lowest accretion history.
\item We have shown further evidence for non-uniform decay of discs in
  SFRs, although new comparisons must be made using consistent
  disc fraction indicators and mass ranges.
\end{enumerate}

\section*{ACKNOWLEDGMENTS}
NJM is funded by a UK Particle Physics and Astronomy Research Council
(PPARC) studentship.
\label{lastpage}

\bsp
\bibliographystyle{mn2e}
\bibliography{references}

\begin{thebibliography}{}

\bibitem[\protect\citeauthoryear{{Ballesteros-Paredes} \&
  {Hartmann}}{{Ballesteros-Paredes} \& {Hartmann}}{2007}]{2007RMxAA..43..123B}
{Ballesteros-Paredes} J.,  {Hartmann} L.,  2007, Revista Mexicana de Astronomia
  y Astrofisica, 43, 123

\bibitem[\protect\citeauthoryear{{Balona} \& {Laney}}{{Balona} \&
  {Laney}}{1996}]{1996MNRAS.281.1341B}
{Balona} L.~A.,  {Laney} C.~D.,  1996, \mnras, 281, 1341

\bibitem[\protect\citeauthoryear{{Barrado y Navascu{\'e}s}, {Stauffer},
  {Morales-Calder{\'o}n}, {Bayo}, {Fazzio}, {Megeath}, {Allen}, {Hartmann} \&
  {Calvet}}{{Barrado y Navascu{\'e}s} et~al.}{2007}]{2007ApJ...664..481B}
{Barrado y Navascu{\'e}s} D.,  {Stauffer} J.~R.,  {Morales-Calder{\'o}n} M.,
  {Bayo} A.,  {Fazzio} G.,  {Megeath} T.,  {Allen} L.,  {Hartmann} L.~W.,
  {Calvet} N.,  2007, \apj, 664, 481

\bibitem[\protect\citeauthoryear{{Bessell}}{{Bessell}}{1979}]{1979PASP...91..5%
89B}
{Bessell} M.~S.,  1979, \pasp, 91, 589

\bibitem[\protect\citeauthoryear{{Bessell}}{{Bessell}}{1990}]{1990PASP..102.11%
81B}
{Bessell} M.~S.,  1990, \pasp, 102, 1181

\bibitem[\protect\citeauthoryear{{Bessell}}{{Bessell}}{2000}]{2000PASP..112..9%
61B}
{Bessell} M.~S.,  2000, \pasp, 112, 961

\bibitem[\protect\citeauthoryear{{Bessell}, {Castelli} \& {Plez}}{{Bessell}
  et~al.}{1998}]{1998A&A...333..231B}
{Bessell} M.~S.,  {Castelli} F.,    {Plez} B.,  1998, \aap, 333, 231

\bibitem[\protect\citeauthoryear{{Blaauw}, {Hiltner} \& {Johnson}}{{Blaauw}
  et~al.}{1959}]{1959ApJ...130...69B}
{Blaauw} A.,  {Hiltner} W.~A.,    {Johnson} H.~L.,  1959, \apj, 130, 69

\bibitem[\protect\citeauthoryear{{Bonatto}, {Bica} \& {Girardi}}{{Bonatto}
  et~al.}{2004}]{2004A&A...415..571B}
{Bonatto} C.,  {Bica} E.,    {Girardi} L.,  2004, \aap, 415, 571

\bibitem[\protect\citeauthoryear{{Brown}, {de Geus} \& {de Zeeuw}}{{Brown}
  et~al.}{1994}]{1994A&A...289..101B}
{Brown} A.~G.~A.,  {de Geus} E.~J.,    {de Zeeuw} P.~T.,  1994, \aap, 289, 101

\bibitem[\protect\citeauthoryear{{Burningham}, {Naylor}, {Littlefair} \&
  {Jeffries}}{{Burningham} et~al.}{2005}]{2005MNRAS.363.1389B}
{Burningham} B.,  {Naylor} T.,  {Littlefair} S.~P.,    {Jeffries} R.~D.,  2005,
  \mnras, 363, 1389

\bibitem[\protect\citeauthoryear{{Buser} \& {Kurucz}}{{Buser} \&
  {Kurucz}}{1978}]{1978A&A....70..555B}
{Buser} R.,  {Kurucz} R.~L.,  1978, \aap, 70, 555

\bibitem[\protect\citeauthoryear{{Caballero}}{{Caballero}}{2007}]{Caballero.20%
07.}
{Caballero} J.~A.,  2007, \aap, 466, 917

\bibitem[\protect\citeauthoryear{{Cardelli}, {Clayton} \& {Mathis}}{{Cardelli}
  et~al.}{1989}]{1989ApJ...345..245C}
{Cardelli} J.~A.,  {Clayton} G.~C.,    {Mathis} J.~S.,  1989, \apj, 345, 245

\bibitem[\protect\citeauthoryear{{Castelli}, {Gratton} \& {Kurucz}}{{Castelli}
  et~al.}{1997}]{1997A&A...318..841C}
{Castelli} F.,  {Gratton} R.~G.,    {Kurucz} R.~L.,  1997, \aap, 318, 841

\bibitem[\protect\citeauthoryear{{Castelli} \& {Kurucz}}{{Castelli} \&
  {Kurucz}}{2004}]{2004astro.ph..5087C}
{Castelli} F.,  {Kurucz} R.~L.,  2004, ArXiv Astrophysics e-prints

\bibitem[\protect\citeauthoryear{{Claria}}{{Claria}}{1982}]{Claria.1982.}
{Claria} J.~J.,  1982, \aaps, 47, 323

\bibitem[\protect\citeauthoryear{{Currie}, {Balog}, {Kenyon}, {Rieke}, {Prato},
  {Young}, {Muzerolle}, {Clemens}, {Buie}, {Sarcia}, {Grabu}, {Tollestrup},
  {Taylor}, {Dunham} \& {Mace}}{{Currie} et~al.}{2007}]{2007ApJ...659..599C}
{Currie} T.,  {Balog} Z.,  {Kenyon} S.~J.,  {Rieke} G.,  {Prato} L.,  {Young}
  E.~T.,  {Muzerolle} J.,  {Clemens} D.~P.,  {Buie} M.,  {Sarcia} D.,  {Grabu}
  A.,  {Tollestrup} E.~V.,  {Taylor} B.,  {Dunham} E.,    {Mace} G.,  2007,
  \apj, 659, 599

\bibitem[\protect\citeauthoryear{{Dahm} \& {Hillenbrand}}{{Dahm} \&
  {Hillenbrand}}{2007}]{2007AJ....133.2072D}
{Dahm} S.~E.,  {Hillenbrand} L.~A.,  2007, \aj, 133, 2072

\bibitem[\protect\citeauthoryear{{Flower}}{{Flower}}{1996}]{1996ApJ...469..355%
F}
{Flower} P.~J.,  1996, \apj, 469, 355

\bibitem[\protect\citeauthoryear{{Garrison}}{{Garrison}}{1970}]{1970AJ.....75.%
1001G}
{Garrison} R.~F.,  1970, \aj, 75, 1001

\bibitem[\protect\citeauthoryear{{Genzel}, {Reid}, {Moran} \&
  {Downes}}{{Genzel} et~al.}{1981}]{1981ApJ...244..884G}
{Genzel} R.,  {Reid} M.~J.,  {Moran} J.~M.,    {Downes} D.,  1981, \apj, 244,
  884

\bibitem[\protect\citeauthoryear{{Girardi}, {Bertelli}, {Bressan}, {Chiosi},
  {Groenewegen}, {Marigo}, {Salasnich} \& {Weiss}}{{Girardi}
  et~al.}{2002}]{2002A&A...391..195G}
{Girardi} L.,  {Bertelli} G.,  {Bressan} A.,  {Chiosi} C.,  {Groenewegen}
  M.~A.~T.,  {Marigo} P.,  {Salasnich} B.,    {Weiss} A.,  2002, \aap, 391, 195

\bibitem[\protect\citeauthoryear{{Haisch} Jr., {Lada} \& {Lada}}{{Haisch}
  et~al.}{2001}]{2001ApJ...553L.153H}
{Haisch} Jr. K.~E.,  {Lada} E.~A.,    {Lada} C.~J.,  2001, \apjl, 553, L153

\bibitem[\protect\citeauthoryear{{Hensberge}, {Pavlovski} \&
  {Verschueren}}{{Hensberge} et~al.}{2000}]{2000A&A...358..553H}
{Hensberge} H.,  {Pavlovski} K.,    {Verschueren} W.,  2000, \aap, 358, 553

\bibitem[\protect\citeauthoryear{{Hern{\'a}ndez}, {Hartmann}, {Megeath},
  {Gutermuth}, {Muzerolle}, {Calvet}, {Vivas}, {Brice{\~n}o}, {Allen},
  {Stauffer}, {Young} \& {Fazio}}{{Hern{\'a}ndez}
  et~al.}{2007}]{2007ApJ...662.1067H}
{Hern{\'a}ndez} J.,  {Hartmann} L.,  {Megeath} T.,  {Gutermuth} R.,
  {Muzerolle} J.,  {Calvet} N.,  {Vivas} A.~K.,  {Brice{\~n}o} C.,  {Allen} L.,
   {Stauffer} J.,  {Young} E.,    {Fazio} G.,  2007, \apj, 662, 1067

\bibitem[\protect\citeauthoryear{{Hillenbrand}}{{Hillenbrand}}{1997}]{1997AJ..%
..113.1733H}
{Hillenbrand} L.~A.,  1997, \aj, 113, 1733

\bibitem[\protect\citeauthoryear{{James}, {Melo}, {Santos} \&
  {Bouvier}}{{James} et~al.}{2006}]{2006A&A...446..971J}
{James} D.~J.,  {Melo} C.,  {Santos} N.~C.,    {Bouvier} J.,  2006, \aap, 446,
  971

\bibitem[\protect\citeauthoryear{{Jeffries}}{{Jeffries}}{2007a}]{2007MNRAS.376%
.1109J}
{Jeffries} R.~D.,  2007a, \mnras, 376, 1109

\bibitem[\protect\citeauthoryear{{Jeffries}}{{Jeffries}}{2007b}]{2007arXiv0707%
.4641J}
{Jeffries} R.~D.,  2007b, ArXiv e-prints, 707

\bibitem[\protect\citeauthoryear{{Jeffries}, {Oliveira}, {Naylor}, {Mayne} \&
  {Littlefair}}{{Jeffries} et~al.}{2007}]{2007MNRAS.376..580J}
{Jeffries} R.~D.,  {Oliveira} J.~M.,  {Naylor} T.,  {Mayne} N.~J.,
  {Littlefair} S.~P.,  2007, \mnras, 376, 580

\bibitem[\protect\citeauthoryear{{Johnson} \& {Morgan}}{{Johnson} \&
  {Morgan}}{1953}]{1953ApJ...117..313J}
{Johnson} H.~L.,  {Morgan} W.~W.,  1953, \apj, 117, 313

\bibitem[\protect\citeauthoryear{{Keller}, {Grebel}, {Miller} \&
  {Yoss}}{{Keller} et~al.}{2001}]{2001AJ....122..248K}
{Keller} S.~C.,  {Grebel} E.~K.,  {Miller} G.~J.,    {Yoss} K.~M.,  2001, \aj,
  122, 248

\bibitem[\protect\citeauthoryear{{Kraus}, {Balega}, {Berger}, {Hofmann},
  {Millan-Gabet}, {Monnier}, {Ohnaka}, {Pedretti}, {Preibisch}, {Schertl},
  {Schloerb}, {Traub} \& {Weigelt}}{{Kraus} et~al.}{2007}]{2007A&A...466..649K}
{Kraus} S.,  {Balega} Y.~Y.,  {Berger} J.-P.,  {Hofmann} K.-H.,  {Millan-Gabet}
  R.,  {Monnier} J.~D.,  {Ohnaka} K.,  {Pedretti} E.,  {Preibisch} T.,
  {Schertl} D.,  {Schloerb} F.~P.,  {Traub} W.~A.,    {Weigelt} G.,  2007,
  \aap, 466, 649

\bibitem[\protect\citeauthoryear{{Krumholz} \& {Tan}}{{Krumholz} \&
  {Tan}}{2007}]{2007ApJ...654..304K}
{Krumholz} M.~R.,  {Tan} J.~C.,  2007, \apj, 654, 304

\bibitem[\protect\citeauthoryear{{Lejeune}, {Cuisinier} \& {Buser}}{{Lejeune}
  et~al.}{1998}]{1998A&AS..130...65L}
{Lejeune} T.,  {Cuisinier} F.,    {Buser} R.,  1998, \aaps, 130, 65

\bibitem[\protect\citeauthoryear{{Lejeune} \& {Schaerer}}{{Lejeune} \&
  {Schaerer}}{2001}]{2001A&A...366..538L}
{Lejeune} T.,  {Schaerer} D.,  2001, \aap, 366, 538

\bibitem[\protect\citeauthoryear{{Lyo}, {Lawson}, {Mamajek}, {Feigelson},
  {Sung} \& {Crause}}{{Lyo} et~al.}{2003}]{2003MNRAS.338..616L}
{Lyo} A.-R.,  {Lawson} W.~A.,  {Mamajek} E.~E.,  {Feigelson} E.~D.,  {Sung}
  E.-C.,    {Crause} L.~A.,  2003, \mnras, 338, 616

\bibitem[\protect\citeauthoryear{{Mathis}}{{Mathis}}{1990}]{1990ARA&A..28...37%
M}
{Mathis} J.~S.,  1990, \araa, 28, 37

\bibitem[\protect\citeauthoryear{{Mayne}, {Naylor}, {Littlefair}, {Saunders} \&
  {Jeffries}}{{Mayne} et~al.}{2007}]{2007MNRAS.375.1220M}
{Mayne} N.~J.,  {Naylor} T.,  {Littlefair} S.~P.,  {Saunders} E.~S.,
  {Jeffries} R.~D.,  2007, \mnras, 375, 1220

\bibitem[\protect\citeauthoryear{{Mendoza V.} \& {Gomez}}{{Mendoza V.} \&
  {Gomez}}{1980}]{1980MNRAS.190..623M}
{Mendoza V.} E.~E.,  {Gomez} T.,  1980, \mnras, 190, 623

\bibitem[\protect\citeauthoryear{{Murdin} \& {Penston}}{{Murdin} \&
  {Penston}}{1977}]{1977MNRAS.181..657M}
{Murdin} P.,  {Penston} M.~V.,  1977, \mnras, 181, 657

\bibitem[\protect\citeauthoryear{{Naylor} \& {Jeffries}}{{Naylor} \&
  {Jeffries}}{2006}]{2006MNRAS.373.1251N}
{Naylor} T.,  {Jeffries} R.~D.,  2006, \mnras, 373, 1251

\bibitem[\protect\citeauthoryear{{Naylor}, {Totten}, {Jeffries}, {Pozzo},
  {Devey} \& {Thompson}}{{Naylor} et~al.}{2002}]{2002MNRAS.335..291N}
{Naylor} T.,  {Totten} E.~J.,  {Jeffries} R.~D.,  {Pozzo} M.,  {Devey} C.~R.,
   {Thompson} S.~A.,  2002, \mnras, 335, 291

\bibitem[\protect\citeauthoryear{{Palla}, {Randich}, {Flaccomio} \&
  {Pallavicini}}{{Palla} et~al.}{2005}]{2005ApJ...626L..49P}
{Palla} F.,  {Randich} S.,  {Flaccomio} E.,    {Pallavicini} R.,  2005, \apjl,
  626, L49

\bibitem[\protect\citeauthoryear{{Park} \& {Sung}}{{Park} \&
  {Sung}}{2002}]{2002AJ....123..892P}
{Park} B.-G.,  {Sung} H.,  2002, \aj, 123, 892

\bibitem[\protect\citeauthoryear{{Perez}, {The} \& {Westerlund}}{{Perez}
  et~al.}{1987}]{1987PASP...99.1050P}
{Perez} M.~R.,  {The} P.~S.,    {Westerlund} B.~E.,  1987, \pasp, 99, 1050

\bibitem[\protect\citeauthoryear{{Pinsonneault}, {Terndrup}, {Hanson} \&
  {Stauffer}}{{Pinsonneault} et~al.}{2004}]{2004ApJ...600..946P}
{Pinsonneault} M.~H.,  {Terndrup} D.~M.,  {Hanson} R.~B.,    {Stauffer} J.~R.,
  2004, \apj, 600, 946

\bibitem[\protect\citeauthoryear{{Prisinzano}, {Damiani}, {Micela} \&
  {Pillitteri}}{{Prisinzano} et~al.}{2007}]{2007A&A...462..123P}
{Prisinzano} L.,  {Damiani} F.,  {Micela} G.,    {Pillitteri} I.,  2007, \aap,
  462, 123

\bibitem[\protect\citeauthoryear{{Robichon}, {Arenou}, {Mermilliod} \&
  {Turon}}{{Robichon} et~al.}{1999}]{1999A&A...345..471R}
{Robichon} N.,  {Arenou} F.,  {Mermilliod} J.-C.,    {Turon} C.,  1999, \aap,
  345, 471

\bibitem[\protect\citeauthoryear{{Sandstrom}, {Peek}, {Bower}, {Bolatto} \&
  {Plambeck}}{{Sandstrom} et~al.}{2007}]{2007arXiv0706.2361S}
{Sandstrom} K.~M.,  {Peek} J.~E.~G.,  {Bower} G.~C.,  {Bolatto} A.~D.,
  {Plambeck} R.~L.,  2007, ArXiv e-prints, 706

\bibitem[\protect\citeauthoryear{{Sanner}, {Altmann}, {Brunzendorf} \&
  {Geffert}}{{Sanner} et~al.}{2000}]{2000A&A...357..471S}
{Sanner} J.,  {Altmann} M.,  {Brunzendorf} J.,    {Geffert} M.,  2000, \aap,
  357, 471

\bibitem[\protect\citeauthoryear{{Sicilia-Aguilar}, {Hartmann},
  {Hern{\'a}ndez}, {Brice{\~n}o} \& {Calvet}}{{Sicilia-Aguilar}
  et~al.}{2005}]{2005AJ....130..188S}
{Sicilia-Aguilar} A.,  {Hartmann} L.~W.,  {Hern{\'a}ndez} J.,  {Brice{\~n}o}
  C.,    {Calvet} N.,  2005, \aj, 130, 188

\bibitem[\protect\citeauthoryear{{Siess}, {Dufour} \& {Forestini}}{{Siess}
  et~al.}{2000}]{2000A&A...358..593S}
{Siess} L.,  {Dufour} E.,    {Forestini} M.,  2000, \aap, 358, 593

\bibitem[\protect\citeauthoryear{{Siess}, {Forestini} \& {Bertout}}{{Siess}
  et~al.}{1999}]{1999A&A...342..480S}
{Siess} L.,  {Forestini} M.,    {Bertout} C.,  1999, \aap, 342, 480

\bibitem[\protect\citeauthoryear{{Slesnick}, {Hillenbrand} \&
  {Carpenter}}{{Slesnick} et~al.}{2004}]{2004ApJ...610.1045S}
{Slesnick} C.~L.,  {Hillenbrand} L.~A.,    {Carpenter} J.~M.,  2004, \apj, 610,
  1045

\bibitem[\protect\citeauthoryear{{Slesnick}, {Hillenbrand} \&
  {Massey}}{{Slesnick} et~al.}{2002}]{2002ApJ...576..880S}
{Slesnick} C.~L.,  {Hillenbrand} L.~A.,    {Massey} P.,  2002, \apj, 576, 880

\bibitem[\protect\citeauthoryear{{Southworth} \& {Clausen}}{{Southworth} \&
  {Clausen}}{2007}]{2007A&A...461.1077S}
{Southworth} J.,  {Clausen} J.~V.,  2007, \aap, 461, 1077

\bibitem[\protect\citeauthoryear{{Southworth}, {Maxted} \&
  {Smalley}}{{Southworth} et~al.}{2004a}]{2004MNRAS.349..547S}
{Southworth} J.,  {Maxted} P.~F.~L.,    {Smalley} B.,  2004a, \mnras, 349, 547

\bibitem[\protect\citeauthoryear{{Southworth}, {Maxted} \&
  {Smalley}}{{Southworth} et~al.}{2004b}]{2004MNRAS.351.1277S}
{Southworth} J.,  {Maxted} P.~F.~L.,    {Smalley} B.,  2004b, \mnras, 351, 1277

\bibitem[\protect\citeauthoryear{{Southworth}, {Zucker}, {Maxted} \&
  {Smalley}}{{Southworth} et~al.}{2004}]{2004MNRAS.355..986S}
{Southworth} J.,  {Zucker} S.,  {Maxted} P.~F.~L.,    {Smalley} B.,  2004,
  \mnras, 355, 986

\bibitem[\protect\citeauthoryear{{Stolte}, {Brandner}, {Brandl}, {Zinnecker} \&
  {Grebel}}{{Stolte} et~al.}{2004}]{2004AJ....128..765S}
{Stolte} A.,  {Brandner} W.,  {Brandl} B.,  {Zinnecker} H.,    {Grebel} E.~K.,
  2004, \aj, 128, 765

\bibitem[\protect\citeauthoryear{{Sung}, {Chun} \& {Bessell}}{{Sung}
  et~al.}{2000}]{2000AJ....120..333S}
{Sung} H.,  {Chun} M.-Y.,    {Bessell} M.~S.,  2000, \aj, 120, 333

\bibitem[\protect\citeauthoryear{{Tout}, {Livio} \& {Bonnell}}{{Tout}
  et~al.}{1999}]{1999MNRAS.310..360T}
{Tout} C.~A.,  {Livio} M.,    {Bonnell} I.~A.,  1999, \mnras, 310, 360

\bibitem[\protect\citeauthoryear{{Westera}, {Lejeune} \& {Buser}}{{Westera}
  et~al.}{1999}]{1999ASPC..192..203W}
{Westera} P.,  {Lejeune} T.,    {Buser} R.,  1999, in {Hubeny} I.,  {Heap} S.,
   {Cornett} R.,  eds, Spectrophotometric Dating of Stars and Galaxies Vol.~192
  of Astronomical Society of the Pacific Conference Series, {Metallicity
  calibration of theoretical stellar SEDs using UBVRIJHKL photometry of
  globular clusters}.
pp 203--+

\bibitem[\protect\citeauthoryear{{Young}, {Lada}, {Teixeira}, {Muzerolle},
  {Muench}, {Stauffer}, {Beichman}, {Rieke}, {Hines}, {Su}, {Engelbracht},
  {Gordon}, {Misselt}, {Morrison}, {Stansberry} \& {Kelly}}{{Young}
  et~al.}{2004}]{2004ApJS..154..428Y}
{Young} E.~T.,  {Lada} C.~J.,  {Teixeira} P.,  {Muzerolle} J.,  {Muench} A.,
  {Stauffer} J.,  {Beichman} C.~A.,  {Rieke} G.~H.,  {Hines} D.~C.,  {Su}
  K.~Y.~L.,  {Engelbracht} C.~W.,  {Gordon} K.~D.,  {Misselt} K.,  {Morrison}
  J.,  {Stansberry} J.,    {Kelly} D.,  2004, \apjs, 154, 428

\end{thebibliography}

\end{document}